\documentclass{ws-ijmpa}
\usepackage[super,compress]{cite}

\usepackage{amsmath}
\usepackage{amssymb}
\usepackage{xr}
\usepackage{graphicx}
\usepackage{mathrsfs}
\usepackage{bm}
\usepackage[hyperref]{xcolor} %
\definecolor{darkgreen}{rgb}{0,0.4,0} %
\definecolor{darkred}{rgb}{0.55,0,0} %
\definecolor{navy}{rgb}{0,0,0.55} %
\usepackage[breaklinks=true,colorlinks=true,citecolor=darkgreen,linkcolor=black,urlcolor=navy]{hyperref} %

\newcommand{\chap}[1]{}
\newcommand{\notchap}[1]{#1}
\newcommand{\chapeight}[1]{}

\newcommand{\prm}[1]{\protect{$#1$}}

\newcommand{\sol}[1]{}

\newcommand{\bmh}{{\bm{\mathrm{H}}}}

\newcommand{\tc}{{\tilde{c}}}

\newcommand{\tW}{{\tilde{W}}}

\newcommand{\mrm}[1]{\mathrm{#1}}

\newcommand{\mbb}[1]{\mathbb{#1}}
\newcommand{\mc}[1]{\mathcal{#1}}

\newcommand{\bgrid}{\left(\begin{array}{rrr}}
\newcommand{\egrid}{\end{array}\right)}
\newcommand{\bgridt}{\left(\begin{array}{rr}}
\newcommand{\egridt}{\end{array}\right)}
\newcommand{\bgridtt}{\left[\begin{array}{rr}}
\newcommand{\egridtt}{\end{array}\right]}
\newcommand{\eref}[1]{(\ref{#1})}
\newcommand{\Eref}[1]{Eq.~(\ref{#1})}

\newcommand{\Erefr}[2]{Eqs.~(\ref{#1})--(\ref{#2})}

\newcommand{\Erefs}[2]{Eqs.~(\ref{#1}) and~(\ref{#2})}

\newcommand{\erefs}[2]{(\ref{#1}) and~(\ref{#2})}

\newcommand{\cref}[1]{Chapter~\ref{#1}}
\newcommand{\Cref}[1]{Chapter~\ref{#1}}

\newcommand{\crefr}[2]{Chapters~\ref{#1}--\ref{#2}}

\newcommand{\fref}[1]{Fig.~\ref{#1}}

\newcommand{\sref}[1]{Sec.~\ref{#1}}

\newcommand{\Sref}[1]{Section~\ref{#1}}
\newcommand{\srefs}[2]{Secs.~\ref{#1} and~\ref{#2}}

\newcommand{\srefr}[2]{Secs.~\ref{#1}--\ref{#2}}
\newcommand{\aref}[1]{Appendix~\ref{#1}}
\newcommand{\Aref}[1]{\ref{#1}}

\newcommand{\tref}[1]{Table~\ref{#1}}
\newcommand{\bsm}{\bar\sigma^\mu}
\newcommand{\bsmm}{\bar\sigma_\mu}
\newcommand{\bsn}{\bar\sigma^\nu}

\newcommand{\nn}{\nonumber}
\newcommand{\rmd}{\mathrm{d}}
\newcommand{\rmi}{\mathrm{i}}%

\newcommand{\Tr}{\mrm{Tr}\,}

\newcommand{\rcite}[1]{Ref.~\citen{#1}}
\newcommand{\rcitet}[1]{Refs.~\citen{#1}}
\newcommand{\prcite}[1]{Ref.~\protect{\citen{#1}}}

\newcommand{\rcites}[2]{Refs.~\citen{#1} and~\citen{#2}}

\newcommand{\p}[1]{\phantom{#1}}

\newcommand{\OO}[1]{\mrm{O}\left(#1\right)}
\newcommand{\OOO}[1]{\mrm{O}\left[#1\right]}
\newcommand{\OOOO}[1]{\mrm{O}\left\{#1\right\}}

\newcommand{\eV}{\mrm{eV}}

\newcommand{\MeV}{\mrm{M}\eV}
\newcommand{\GeV}{\mrm{G}\eV}
\newcommand{\TeV}{\mrm{T}\eV}
\newcommand{\SU}[1]{\mrm{SU}(#1)}
\newcommand{\su}[1]{\mrm{su}(#1)}

\newcommand{\U}[1]{\mrm{U}(#1)}

\newcommand{\peref}[1]{\protect{(\ref{#1})}}

\newcommand{\pAref}[1]{\protect{\ref{#1}}}
\newcommand{\pcref}[1]{\protect{Chapter~\ref{#1}}}
\newcommand{\pcite}[1]{\protect{\cite{#1}}}

\newcommand{\GL}[2]{\mrm{GL}(#1,\mbb{#2})}

\newcommand{\gl}[2]{\mrm{gl}(#1,\mbb{#2})}
\newcommand{\SL}[2]{\mrm{SL}(#1,\mbb{#2})}
\renewcommand{\sl}[2]{\mrm{sl}(#1,\mbb{#2})}

\renewcommand{\bar}[1]{\overline{#1}}

\newcommand{\preon}{\psi}

\newcommand{\ILO}[1]{\mrm{O}(#1)}

\newcommand{\RM}{{{\mbb{R}^{1,3}}}}

\newcommand{\Cw}[1]{{{\mbb{C}^{\wedge #1}}}}

\newcommand{\tagref}[1]{{\tag{\ref{#1}}}}

\newcommand{\taue}{e}

\newcommand{\quietGversion}[1]{}
\newcommand{\onlyinsummary}[1]{}

\newcommand{\midrule}{\colrule}

\newcommand{\achapnotchap}[2]{(\notchap{\rcite{#2},~}\aref{#1})}
\newcommand{\schapnotchap}[2]{(\notchap{\rcite{#2},~}\sref{#1})}
\newcommand{\cchapnotchap}[2]{(\notchap{\rcite{#2},~}\cref{#1})}
\newcommand{\paper}{\chap{chapter}\notchap{paper}}
\newcommand{\draft}[1]{#1} %
\newcommand{\draftone}[1]{#1}
\newcommand{\drafttwo}[1]{#1}

\begin{document}
\markboth{R.~N.~C.~Pfeifer}{Analog predictions for $W$~mass}%

\catchline{}{}{}{}{}

\title{Falsifiable analog model predictions of $W$~mass in CDF~II and ATLAS}

\author{R.~N.~C.~Pfeifer}

\address{Dunedin, Otago, New Zealand\\rncpfeifer@gmail.com}

\maketitle

\begin{history}
January 1, 2024
\end{history}

\chap{This Chapter has been published as:\\RNC Pfeifer, \emph{Falsifiable analog model predictions of $W$~mass in CDF~II and ATLAS},\\\emph{Int. J. Mod. Phys. A}%
, DOI: \href{https://doi.org/10.1142/S0217751X23400043}{10.1142/S0217751X23400043}.\\
Updates as listed in the preface. In particular, Table~\ref{tab:masses} and \Eref{eq:test2} have been updated to reflect improved calculation of $m_H$ as per \sref{sec:scalbosonmass}.\\
Licensed under \href{http://creativecommons.org/licenses/by/4.0/}{CC BY 4.0}.\\~\\}

\begin{abstract}
The CDF~II measurement of $W$~boson mass is at significant ($6.9\,\sigma$) tension with the Standard Model, and moderate ($4.0\,\sigma$) tension with recent results from ATLAS. Any attempt to interpret this as a signature of new physics requires high-precision, robust, falsifiable predictions. The Classical Analogue to the Standard Model In pseudo-Riemannian spacetime (CASMIR) is an analogue model predicting $W$ and $Z$ boson masses of $80.3587(22)~\GeV/c^2$ and $91.1877(35)~\GeV/c^2$ respectively\draftone{. During baryon collisions satisfying $\sqrt{s}< 3.09~\TeV$, CASMIR predicts a color-mediated enhancement of $W$ and $Z$ boson masses, becoming $80.4340(22)~\GeV/c^2$ and $91.1922(35)~\GeV/c^2$ respectively.
The unenhanced masses are consistent with ATLAS data collected at $\sqrt{s}=7~\TeV$, and the enhanced masses are consistent with CDF~II data collected at $\sqrt{s}=1.96~\TeV$. According to CASMIR, operation of the LHC at centre-of-mass energies small compared with $3.09~\TeV$ but large %
enough for $W$~boson formation should permit ATLAS to replicate the result from CDF~II.}

\end{abstract}

\ccode{PACS numbers: 14.70.Fm, 14.60.Ef, 12.60.-i\\~\\This article has been published as:\\RNC Pfeifer, \emph{Falsifiable analog model predictions of $W$~mass in CDF~II and ATLAS},\\\emph{Int. J. Mod. Phys. A}%
, DOI: \href{https://doi.org/10.1142/S0217751X23400043}{10.1142/S0217751X23400043}.\\
Licensed under \href{http://creativecommons.org/licenses/by/4.0/}{CC BY 4.0}. Table~\ref{tab:masses} and \Eref{eq:test2} have been updated to reflect improved calculation of $m_H$ in \rcite{pfeifer2022CASM4}.} %

\section{Introduction}

The Standard Model is undeniably the greatest achievement of modern particle physics, %
displaying unrivaled congruence between theory and experiment~\cite{glashow1961,salam1964,weinberg1967,awramik2004,erler2019,workman2022}. 
However, recent results from Fermilab raise the possibility of detecting physics beyond the Standard Model, with the CDF~II measurement of $W$~boson mass %
showing a statistically significant discrepancy of $6.9\,\sigma$ when compared with the Standard Model prediction~\cite{awramik2004,erler2019,workman2022,aaltonen2022}. 
Similarly, measurement of muon magnetic anomaly in the Muon \mbox{$g-2$} experiment %
shows a discrepancy of \draftone{$5.2\,\sigma$} with respect to the Standard Model value~\cite{aguillard2023a,aoyama2020,keshavarzi2022}.

While tensions between CDF~II and ATLAS are smaller than those between CDF~II and the Standard Model ($4.0\,\sigma$ compared with $6.9\,\sigma$%
)~\cite{aaltonen2022,awramik2004,erler2019,workman2022,the-ATLAS-collaboration2023}, it is highly possible that further data collection at ATLAS might yield a value distinct from CDF~II at the level of $5\,\sigma$ or better, representing a discovery. Of course, most such discoveries are prosaic, %
e.g.{}
\begin{itemize}
\item an unrecognized systematic experimental error, or
\item an incomplete understanding of background events mimicking the $W$~boson signal,
\end{itemize}
and it is far more rare for such a tension to represent an actual novel interaction or process leading to a true difference in the measured values.
For the latter possibility, the standard of proof must therefore be set appropriately high before any such species or interaction may seriously be contemplated.

\draftone{%
A similar situation exists for the muon gyromagnetic anomaly. Although recent data have increased the precision of the Fermilab measurement of muon~\mbox{$g-2$} such that the tension with the 2020 theoretical prediction now exceeds the $5.0\,\sigma$ threshold~\cite{aguillard2023a}, this is tempered by other recent developments:
a lattice gauge calculation of the Hadron Vacuum Polarisation (HVP)~\cite{borsanyi2021} and a measurement of the $e^+e^-\rightarrow \pi^+\pi^-$ cross-section at \mbox{CMD-3}~\cite{ignatov2023}, both of which disagree with previous $e^+e^-$ experiments, with implications for the Standard Model calculation.
Attempts to reconcile these findings with previous results are ongoing~\cite{colangelo2022}, and
for the present the $5.0\,\sigma$ threshold is not yet considered to have been decisively crossed.} %

\draftone{The present \paper{} approaches the unexpected results from CDF~II and Muon~\mbox{$g-2$}} through the application of an analogue model, the Classical Analogue to the Standard Model In pseudo-Riemannian spacetime (CASMIR). This model was originally developed by the author to assist in exploring certain interactions involving the Standard Model in a gravitational field.
By construction it displays high congruence with the Standard Model across almost all experimentally accessible regimes, and %
key Standard Model parameters %
not input into the model are reproduced
to within $0.2\,\sigma$ of their observed values (\Aref{apdx:massrel}). %
Discrepancies between CASMIR and the Standard Model arise when computing the masses of the $W$ and $Z$ bosons, with CASMIR predicting twinned mass eigenvalues for both species.

Remarkably, predictions from CASMIR precisely reproduce the $W$~\draftone{boson} mass values observed in both ATLAS and CDF~II, with precision greater than experiment and tensions of $0.1\,\sigma$ or better. %
This occurs through the presence in CASMIR of eight additional colored $W$~\draftone{boson mass eigenstates which} are predicted to be excluded from the $W$~\draftone{boson} mass \draftone{measurement} at ATLAS, but retained \draftone{at} CDF~II. In CASMIR the conventional $W$~boson has a calculated mass of $80.3587(22)~\GeV/c^2$, and the colored boson \draftone{eigenstates} have masses of $80.4434(22)~\GeV/c^2$. The detected value at CDF~II is the r.m.s.{} mass across the nine $W$-like \draftone{mass eigenstates}, yielding a predicted value of $80.4340(22)~\GeV/c^2$. There are no fitting parameters in this calculation\drafttwo{, and effects on $W$~and $Z$ boson cross-sections are small compared with current experimental precision}.

The value of the muon gyromagnetic anomaly also differs between CASMIR and the Standard Model, with the CASMIR value being $116592071(46)\times 10^{-11}$ compared with $116591810(43)\times 10^{-11}$ in the Standard Model.
This difference arises because 
under a specific set of conditions the geometry of CASMIR allows electroweak bosons to also carry a color charge. Again there are no fitting parameters.

An analogue model is only useful if it forms the basis for robustly (and preferably rapidly) testable, falsifiable predictions, ideally with existing equipment and/or datasets. Based on the analysis presented in this \paper{}, 
\draftone{the enhancement of the $W$~boson mass observed in CDF~II is energy-dependent, and may be replicated in ATLAS by reducing the LHC centre-of-mass energy 
$\sqrt{s}$ to be small compared with $3.09~\TeV$.
}

The energy scales of these predictions all fall firmly within the projected domain of validity of the CASMIR model. If verified, these predictions would provide an intriguing suggestion of novel bosons (or somehow mass-shifted versions of the existing $W$~boson) at $80.4434(22)~\GeV/c^2$. %
If refuted this would be less interesting to the particle physics community at large, but would %
bound the domain of correspondence between CASMIR and the Standard Model at $E\ll m_Wc^2\sim 80~\GeV$ rather than $3.1~\TeV$ as presently anticipated.

\section{Overview: Limits of CASMIR\label{sec:limitsoverview}}

This Section provides an overview of known regimes in which accelerator physics predictions from CASMIR are anticipated to deviate from the Standard Model, with brief discussion of the accessibility of each of these regimes. It provides context for the more in-depth analysis of the accessible discrepancies and resulting predictions which is provided in \srefr{sec:weakmasses}{sec:muong2}.

\subsection{Weak boson masses\label{sec:overviewWeakMasses}}

A key discrepancy between CASMIR and the Standard Model is the prediction in CASMIR of paired eigenvalues for the $W$ and $Z$ boson masses, with the components of each doublet differing by less than $90~\MeV/c^2$. %
No suggestion of such a doublet has been seen in any single accelerator dataset, though it is only recently that the weak boson masses have been measured to sufficient precision to permit discrimination of such closely-spaced values. It is also noted that in CASMIR the upper eigenvalues of each doublet are associated with the presence of a color charge, \draftone{which carries implications for the environments in which these eigenvalues may be observed.} %

In \sref{sec:weakmasses} these properties of CASMIR are assessed for consistency with existing datasets, in particular LEP, ATLAS, and CDF~II. %
On the basis of the available information, and with the exception of the $W$~\draftone{boson} mass measurement in CDF~II, %
existing datasets are not provably able either to detect or to exclude the upper $W$ and $Z$ \draftone{boson} mass eigenvalues predicted by CASMIR.

Prospects for additional near-term tests of CASMIR are discussed, including \draftone{%
prospects for 
reproduction of the CDF~II result in ATLAS.}

\subsection{Muon $g-2$}

As a consequence of differences in the electroweak sector, when the muon gyromagnetic anomaly is measured at the muon ``magic momentum''~\cite{abi2021}, CASMIR predicts a higher value for the muon gyromagnetic anomaly than the Standard Model. While a full-precision re-evaluation of the muon gyromagnetic anomaly using CASMIR is beyond the scope of the present \paper{}, the natures of these changes to the electroweak sector make it straightforward to determine bounds on the result of such a calculation (see \sref{sec:muong2}). The resulting \draftone{value} of $116592071(46)\times 10^{-11}$ shows significantly reduced tension with experimental results~\cite{aguillard2023a} with only a small increase in uncertainty, and this reduction in tension exceeds that which can be attributed to expansion of the confidence interval alone. Consequently CASMIR predicts the muon gyromagnetic anomaly more accurately at the muon magic momentum than the Standard Model, and hence---despite this discrepancy with the Standard Model---measurements of muon~\mbox{$g-2$} do not constrain the applicability of CASMIR.

\subsection{Preferred rest frame and finite cutoff\label{sec:OverviewCutoff}}

An unusual property of CASMIR is the existence of a pseudovacuum state with a preferred rest frame. This pseudovacuum is associated with an energy scale $\mc{E}_\Omega\approx 6.18~\TeV$, and is responsible for the CASMIR analogue of quantum fluctuations in energy and momentum\notchap{~\cite{pfeifer2022CASM1,pfeifer2022CASM4}}. However, the finite value of $\mc{E}_\Omega$ implies a similarly finite UV cutoff of
$\frac{1}{2}\mc{E}_\Omega\approx 3.1~\TeV$ in the isotropy frame of the cosmic microwave background (CMB).\footnote{Energy scales in CASMIR correspond to the inverse of autocorrelation %
scales in the pseudovacuum. Since a field excitation of wavelength \prm{\lambda} exhibits autocorrelation across regions of length \prm{\tfrac{1}{2}\lambda}, a CASMIR energy scale \prm{\mc{E}} corresponds to an excitation energy \prm{E=\tfrac{1}{2}\mc{E}}. In this \paper{}, \prm{\mc{E}} denotes a CASMIR energy scale while \prm{E} denotes the energy of a particle or field.}
As a consequence, processes whose dominant terms involve energy or momentum fluctuations \draftone{that are} large compared with $\frac{1}{2}\mc{E}_\Omega$ when evaluated in the rest frame of the CMB are heavily suppressed in CASMIR. Similarly, the pseudovacuum may cease to appear homogeneous to particles accelerated to energies of $\OO{\frac{1}{2}\mc{E}_\Omega}$ in the rest frame of the CMB.

I am unaware of any practical proposals to test the ability of quantum fluctuations to exceed $3.1~\TeV$, or to use the circulating beamlines of the LHC as a probe for inhomogeneity of the vacuum at high boost. Thus, although the existence of a preferred rest frame in CASMIR represents a significant theoretical break from the Standard Model, as yet it provides no prospects for experimental discrimination between CASMIR and the Standard Model.

\subsection{Higher generation weak bosons}

In addition to the color-mediated weak boson mass doublets discussed in \sref{sec:overviewWeakMasses}, CASMIR also predicts the existence of higher-generation weak bosons, starting with a second-generation $W$~boson at $16.6~\TeV/c^2$ (see \tref{tab:heavyweak}). However, the energy scale of even the lightest of these bosons is presently out of reach of the LHC. 
\begin{table}
\tbl{Masses of the first, second, and third generation uncolored weak bosons in CASMIR.}
{
\chap{\begin{center}}\chapeight{\begin{center}}
\begin{tabular}{cr@{.}lr@{.}lr@{.}l}
\toprule
&\multicolumn{6}{c}{Mass}\\
Boson&\multicolumn{2}{c}{First Generation ($\GeV/c^2$)}&\multicolumn{2}{c}{Second Generation ($\TeV/c^2$)}&\multicolumn{2}{c}{Third Generation ($\TeV/c^2$)}\\
\midrule
$W$&$\qquad\qquad$80&3587(22)&$\qquad\qquad$16&61320(46)&$\qquad\qquad$279&4038(77)\\
$Z$&$\qquad\qquad$91&1877(35)&$\qquad\qquad$18&84673(73)&$\qquad\qquad$316&968(12)\\
$\bmh$&$\qquad\qquad$125&1261(48)&$\qquad\qquad$25&8643(10)&$\qquad\qquad$434&991(17)\\
\botrule
\end{tabular}\label{tab:heavyweak}
\chap{\end{center}}\chapeight{\end{center}}
}
\end{table}%
While these higher-generation bosons may also appear as virtual particles, any contributions to currently-accessible processes from such particles would be heavily suppressed due to the high energies of the higher-generation bosons, and in CASMIR also due to the finite vacuum energy cutoff discussed in \sref{sec:OverviewCutoff}. These bosons display exotic behaviors in CASMIR, with particle generation not behaving as a good quantum number at their interaction vertices. However, no detectable discrepancies are predicted in presently accessible experimental regimes.

\subsection{Dark matter\label{sec:DM}}

CASMIR also describes three species of dark matter, denoted $N_\mu$, $[N_2]_\mu$, and $[N_3]_\mu$\notchap{ in keeping with the %
notation of \rcite{pfeifer2022CASM4}}. These species have inertial masses of $80.3810(22)~\GeV/c^2$, $16.61320(46)~\TeV/c^2$, and $279.4038(77)~\TeV/c^2$ respectively, but break the weak principle of equivalence, with all having a gravitational mass of only $1381.486(37)\,m_e$ \cchapnotchap{ch:gravity}{pfeifer2022CASM4}. %
With their only significant nongravitational interactions being their coupling to the scalar boson, prospects for direct detection of these species are minimal. Since negligible attention has been paid to proposals for dark matter which break the weak principle of equivalence, it is not presently possible to say whether these species are consistent with cosmological observations. They have no impact on the accelerator physics considered in the present \paper{} and are therefore ignored.

\subsection{Neutrino mass and oscillations}

Neutrinos in CASMIR are massless and Weyl, with equilibration between generations arising not as a consequence of neutrino oscillation, but through nonconservation of particle generation at $W$~boson vertices \schapnotchap{sec:neutrinos}{pfeifer2022CASM4}. Consequently there exists a discrepancy in the area of neutrino physics between CASMIR and the Standard Model with massive neutrinos and a seesaw mechanism. Direct detection of neutrino inertial mass would place limitations on the application of CASMIR, as would a lower bound on the length scale over which neutrino oscillations take place. \draftone{N}either of these events have occurred\draftone{. Likewise, CASMIR does not support neutrinoless double beta decay, and neutrinoless double beta decay has also not been observed.}
\draftone{As yet}, 
neutrino physics does 
not place bounds %
on the regime of congruency between CASMIR%
\draftone{, experiment, and} the Standard Model. %

\subsection{\draftone{Additional notes on CASMIR and weak sector constraints}\label{sec:addconstrs}}

\draftone{The weak sector provides ample opportunities to constrain the validity of beyond-Standard-Model constructions. As CASMIR is relatively recently developed, the process of assessment against weak sector constraints is ongoing. However, the following considerations offer reason to expect a high degree of congruency.}

\draftone{In \sref{sec:weakmix} \notchap{of \rcite{pfeifer2022CASM4}} it is established that the effective lepton/$W$ and lepton/$Z$ vector boson interaction terms in the emergent CASMIR Lagrangian show high congruency with those of the Standard Model. Further, the preonic substructure of the fermions permits this conclusion to be generalized to interactions involving quarks, and conserved symmetries (and construction from preons) likewise permit this conclusion to be generalized to three-particle and four-particle vector boson vertices. For fermions and vector bosons, the electroweak sector of the CASMIR model is found to be wholly consistent with the electroweak sector of the Standard Model below the $1\,\sigma$ level, including reproduction of decay rates and branching ratios, up to effects due to colored (chromatic) weak vector bosons discussed in \srefs{sec:weakmasses}{sec:muong2} of the present \paper{}.}

\draftone{For processes involving Higgs bosons, direct assessment of constraints on CASMIR have so far largely been qualitative, and so far have also been consistent with expectations. A few quantitative calculations involving the Higgs boson have been performed, though these %
have no counterpart in the Standard Model. %
They are:
\begin{itemize}
\item Precision calculation of the mass of the Higgs boson \schapnotchap{sec:scalbosmass}{pfeifer2022CASM4}.
\item Relationships between lepton and vector boson masses, including Higgs boson vacuum expectation value contributions to the lepton and vector boson masses and Higgs boson loop corrections to the lepton and vector boson mass interactions (\notchap{\rcite{pfeifer2022CASM4}, }\srefs{sec:VI:bosonmasses}{sec:lepmassint}).
\item Contributions to the calculation of muon~%
\mbox{$g-2$} 
(\sref{sec:muong2} of the present \paper{}).
\end{itemize}
The mass calculations yield lepton, vector boson, and Higgs boson masses %
consistent with observation %
(see \Aref{apdx:massrel}), while the CASMIR muon~\mbox{$g-2$} calculation shows less tension with experiment than the Standard Model (see \sref{sec:muong2}).} %

\draftone{An important test which has not yet been performed is the comparison of predicted Higgs boson production rates with the Standard Model and with observation. In CASMIR, Higgs boson production at energy scales small compared with $3.1~\TeV$ is suppressed by a factor of approximately $1.7\times 10^{-2}$ per vertex, denoted $\big[k^{(e)}_1(\mc{E}_e)N_0\big]^{-2}$ \schapnotchap{sec:scalbosint}{pfeifer2022CASM4}. However, this suppression effect vanishes above the $3.1~\TeV$ threshold. Its relevance or otherwise is as yet unconfirmed in scalar-boson-generating collisions at the LHC, where the beam energy scales are $\sqrt{s}=7~\TeV$ and above, but the energies of the scalar bosons generated in these collisions are substantially more modest.%
\footnote{This is now explored further in \pcref{ch:Higgs}\notchap{ of \rcite{pfeifer2022CASM4}}.}
}

\draftone{Although further testing of CASMIR is necessary, %
the model has so far 
demonstrated good performance against %
electroweak constraints both %
in congruence with the Standard Model and %
in precision calculations extending 
beyond %
the Standard Model.}

\subsection{Summary of limits}

The discrepancies between CASMIR and the Standard Model which are likely to impact upon present experimental results are those which arise from the existence of additional weak sector vector boson \draftone{mass eigenstates} in CASMIR. These additional \draftone{eigenstates correspond to} colored versions of the $W$ and $Z$ bosons, \draftone{having} masses within a hundred $\MeV/c^2$ of their colorless counterparts. 
\draftone{Incorporating the colored mass eigenstates into the calculation of $W$~boson mass yields} a good match for the result from CDF~II, and existence of these \draftone{additional mass eigenstates} also favorably impacts calculation of the muon gyromagnetic anomaly.

In light of these discrepancies, \sref{sec:weakmasses} addresses why current evaluation of the CDF~II dataset is predicted by CASMIR to be sensitive to these novel species while ATLAS is not, and provides \draftone{a prescription} for \draftone{detection of the higher-mass eigenstates in ATLAS}.
\Sref{sec:muong2} then adjusts the calculation of the muon gyromagnetic anomaly to include the CASMIR colored weak boson \draftone{mass eigenstates}. The CASMIR result is consistent with experiment with tension well below $1\,\sigma$, demonstrating that results from the Muon~\mbox{$g-2$} experiment do not constrain the application of CASMIR as an analogue to the Standard Model in the relevant energy regimes.

\section{Weak boson masses\label{sec:weakmasses}}

\subsection{Background}

At present some of the most stringent tests of the Standard Model come from large-scale collider experiments such as CMS~\cite{chatrchyan2012} and ATLAS~\cite{aad2012,aaboud2018,aaboud2018a,the-ATLAS-collaboration2023} at the Large Hadron Collider (LHC), along with ongoing processing of the data from the discontinued CDF~II %
at Tevatron~\cite{aaltonen2022}. %
Such results may be considered independently where they possess some unique merit (for example, extraordinary precision~\cite{aaltonen2022} or uniquely high beam energies~\cite{aaboud2018,aaboud2018a}), or collectively where construction of a global average may amplify statistical power and reduce error~\cite{workman2022}.
Measurement of the mass of the $W$~boson is one of the most challenging
of these tests, as its decay products include an antineutrino \draftone{(for $W^-$ boson decay) or a neutrino (for $W^+$ boson decay)} which is typically lost to detection.
The most precise experimental determination of $W$~boson mass yet performed is that of the CDF collaboration, using 8.8~inverse femtobarns of integrated luminosity collected in proton--antiproton collisions at $1.96~\TeV$ centre-of-mass energy in the CDF~II collector at the Fermilab Tevatron collider~\cite{aaltonen2022}. This data series produced the surprising result of $m_W=80.4335(94)~\GeV/c^2$, in tension with the Standard Model prediction of $80.356(6)~\GeV/c^2$ at $6.9\,\sigma$~\cite{awramik2004,erler2019,workman2022}, the PDG global average of $80.377(12)~\GeV/c^2$ at a level of $3.7\,\sigma$~\cite{workman2022}, and the ATLAS collaboration measurement of $80.360(16)~\GeV/c^2$ at $4.0\,\sigma$~\cite{the-ATLAS-collaboration2023}.\notchap{$\,$}\footnote{This recently-announced unpublished result improves over the previously published measurement of $80.370(19)~\GeV/c^2$\notchap{.}\chap{~}\pcite{aaboud2018,aaboud2018a}\chap{.}} No mechanism has been identified within the Standard Model to explain this discrepancy, so it is not unreasonable to seek an explanation beyond the Standard Model. 

\subsection{CASMIR and chromatic bosons---a brief overview}

This subsection begins in \srefr{sec:particlesCASMIR}{sec:CASMIRmass} with a brief overview of particle construction and mass mechanisms in CASMIR. While not intended as a substitute for the full exposition\chap{ of the preceding chapters},\notchap{\cite{pfeifer2022CASM1,pfeifer2022CASM4,pfeifer2022CASM2,pfeifer2022CASM3}}
it serves to contextualize the subsequent discussion of $W$ and $Z$ boson masses and detection within the CASMIR model.
This is followed in \sref{sec:chromenv} by a discussion of the environments in which chromatic bosons may be detected, before proceeding to specific discussion of the chromatic $W$ and $Z$ bosons in \srefs{sec:chrWbos}{sec:chrZbos} respectively. \drafttwo{\Sref{sec:sxBR} considers the effects of these bosons on cross-sections and branching ratios, and consequently on the rate of $W$~and $Z$ boson signal detection in collider experiments.}

\subsubsection{Particles in CASMIR\label{sec:particlesCASMIR}}

CASMIR is a heirarchical effective field theory model in which all species observed in the Standard Model are constructed from nine charged preons, %
and these preons themselves emerge as an effective low-energy description of the gradients of scalar fields on a higher-dimensional manifold. Although the microscopic origin of the preons is specific to CASMIR, once they are constructed these fields display the familiar structure of a sequence of effective descriptions valid over a descending ladder of energy scales.\footnote{%
There are two principal energy thresholds %
in CASMIR, namely:
\prm{\mc{E}_\preon}, below which preons condense into fermions and bosons and the pseudovacuum admits a (non-homogeneous) continuum description, and \prm{\mc{E}_0}, below which the pseudovacuum appears homogeneous and chromatic bosons and the pseudovacuum structure can no longer be observed. Both are measured in the isotropy frame of the pseudovacuum.
}
This sort of multiscale description is ubiquitous throughout particle physics---for example, hadrons are (quasi)particles composed from collections of quarks, atomic nuclei are composed in turn from hadrons, and yet other quasiparticles such as phonons may emerge from the collective behaviors of atoms~\cite{serot1997,srivastava1990}. 

In CASMIR, all species observed in the Standard Model are quasiparticles constructed from preons carrying charges with respect to a $\GL{9}{R}\otimes\SL{2}{C}$ symmetry. The $\GL{9}{R}$ subgroup behaves as an internal local symmetry, while $\SL{2}{C}$ is the double cover of the local space--time symmetry %
of $\RM$.
There are nine preons, and these are acted on by the eighty-one bosons of the internal symmetry group $\GL{9}{R}$. However, in the presence of $\SL{2}{C}$,
symmetry group $\GL{9}{R}$ admits the decompositions~\schapnotchap{sec:symC18}{pfeifer2022CASM3}
\begin{equation}
\begin{split}
\GL{9}{R}\cong\,&\SU{9}\oplus\GL{1}{R}\\
\cong\,&[\SU{3}_A\oplus\GL{1}{R}_A]%
\otimes[\SU{3}_C\oplus\GL{1}{R}_C]
\end{split}
\end{equation}
where the groups $\SU{3}_B$ and $\GL{1}{R}_B$ for $B\in\{A,C\}$ may alternatively be understood as dimension-8 and dimension-1 representations of symmetry $\SU{3}_B$ respectively. 
In the physical gauge the $\SU{3}_A$ symmetry is broken to $\SU{2}\otimes\U{1}$, with radiative corrections adjusting the coupling constants to reproduce the electroweak sector of the Standard Model\chap{~(\crefr{ch:SM}{ch:detail})}, up to the inclusion of the additional colored weak vector bosons.\notchap{\cite{pfeifer2022CASM3,pfeifer2022CASM4}} The group $\GL{1}{R}_A$ thus corresponds to a trivial representation of the electroweak symmetry $\SU{2}\otimes\U{1}$, and $\GL{1}{R}_C$ to a trivial representation of the color symmetry $\SU{3}_C$. 

Under this decomposition the nine preons each carry two labels, corresponding to charge on the $A$ and $C$~sectors, and color-neutral triplets of identical $A$-charge correspond (up to a normalization factor) to the three families of leptonic Weyl spinors,\notchap{\footnote{This \paper{} uses Weyl spinor notation for fermions, as in e.g.~\protect{\rcite{wess1981}}.}}
\begin{equation*}
\begin{tabular}{c|c}
Spinor triplet~&~Lepton family\\\hline\hline
$\psi^{1r}\psi^{1g}\psi^{1b}$~&~$\{\bar{e}_R,\bar{\mu}_R,\bar{\tau}_R\}$\\
$\psi^{2r}\psi^{2g}\psi^{2b}$~&~$\{e_L,\mu_L,\tau_L\}$\\ 
$\psi^{3r}\psi^{3g}\psi^{3b}$~&~$\{\nu_e,\nu_\mu,\nu_\tau\}$,
\end{tabular}
\end{equation*}
with the relative phases of the three components determining the mass-matrix eigenvalue and thus particle generation of a given preon triplet~\cchapnotchap{ch:fermion}{pfeifer2022CASM4}. Quarks correspond to triplets with mismatched $A$-charge, where these are permitted by %
gauge, and have a residual color due to variations in the strength of confinement of the differing preons%
~(\notchap{\rcite{pfeifer2022CASM3}, }\draftone{\sref{sec:strongint}}).

Coupling of fermions to the boson field $N_\mu$ associated with the trivial subgroup $\GL{1}{R}_A\otimes\GL{1}{R}_C$ is suppressed over length scales large compared with $\mc{L}_0$ by gauge choices~\erefs{eq:U1gauge}{eq:GL1RNgauge} of \chap{\cref{ch:SM}}\notchap{\rcite{pfeifer2022CASM3}}. However, fermions may continue to couple to bosons carrying trivial charge on one of the two subgroups, giving rise to the pure $A$-sector and pure $C$-sector bosons, or carrying trivial charge on neither of the two subgroups, corresponding to bosons carrying both $A$- and $C$-sector charges.

With a boson being composed of one preon and one antipreon, it might be anticipated that the off-diagonal construction $\frac{1}{\sqrt{3}}\bar\nu_e^{\dot c}\bsmm\delta_{\dot cc} e^c_L$ intended to function as the $W$~boson would only change the species of one preon within a fermion triplet.
However, taking one fermion as inbound and one as outbound, a freedom of choice of \draftone{coordinate} frame on the $\GL{9}{R}$ bundle may be fixed by requiring that the $A$-sector bosons act on the fermions (as well as the preons) as a representation of $\su{3}_A$ \schapnotchap{sec:EWint_Wintdetail}{pfeifer2022CASM3}. %
The resulting frame exhibits a discontinuity whose precise location admits some flexibility, with any observer being free to choose its local effects simultaneous with the interaction in their own reference frame, as a result of which
the $A$-charges of the other two members of the triplet are also transformed in association with the interaction. Thus the action of the $W$~boson on the lepton sector is through vertices having the expected form
\begin{equation}
W_\mu \bar e_L\bsm \nu_e\quad\textrm{and h.c.}
\end{equation} 
and similar for higher generations.

If all of the bosons acting purely on one sector are now made explicit,
including those which vanish in the physical gauge, then this yields nine bosons on each of the two sectors. %
It then suffices over macroscopic scales (e.g.~approaching the classical limit, or in many-particle scattering processes such as mass generation) %
to model the eighty-one bosons of $\gl{9}{R}$ as pairwise products of these sets, e.g.{}
\begin{equation}
\bm{\phi}_\mu^{\dot a\dot c a c}(x) \sim \int\rmd^4y~a^{\dot a a}_\nu(y)\bar\Psi(y)\bsn\Psi(x) c^{\dot cc}_\mu(x),\label{eq:compositeSU9}
\end{equation}
where $\Psi(x)$ represents an appropriately-chosen preon triplet.
However, in single-particle processes this representation of the boson sector is misleading, and to accurately model the transport of charge and momentum between fields it is necessary to also take into account %
processes transferring both $A$- and $C$-charge on a single particle. 

Introduce the rescaled Gell-Mann matrices 
\begin{align}
\lambda_1=\frac{1}{\sqrt{2}}\bgrid 0&1&0\\1&0&0\\0&0&0 \egrid \quad&
\lambda_2=\frac{1}{\sqrt{2}}\bgrid 0&-\rmi&0\\\rmi&0&0\\0&0&0 \egrid \quad& %
\lambda_3=\frac{1}{\sqrt{2}}\bgrid 1&0&0\\0&-1&0\\0&0&0 \egrid \nn\\ %
\lambda_4=\frac{1}{\sqrt{2}}\bgrid 0&0&1\\0&0&0\\1&0&0 \egrid \quad& %
\lambda_5=\frac{1}{\sqrt{2}}\bgrid 0&0&-\rmi\\0&0&0\\\rmi&0&0 \egrid \quad&
\lambda_6=\frac{1}{\sqrt{2}}\bgrid 0&0&0\\0&0&1\\0&1&0 \egrid \label{eq:Cbasis}\\
\lambda_7=\frac{1}{\sqrt{2}}\bgrid 0&0&0\\0&0&-\rmi\\0&\rmi&0 \egrid \quad&
\lambda_8=\frac{2}{\sqrt{6}}\bgrid \frac{1}{2}&0&0\\0&\frac{1}{2}&0\\0&0&-1 \egrid \quad& %
\lambda_9=\frac{1}{\sqrt{3}}\bgrid 1&0&0\\0&1&0\\0&0&1 \egrid, \nn%
\end{align}
which satisfy $\Tr{({\lambda_\tc}^2)}=1$ for all $\tc\in\{1,\ldots,9\}$,
and recognize that the achromatic (i.e.~colorless) $W$~boson is associated with representation $\lambda_9$ on the $C$~sector, denoted $\lambda^C_9$. This is then necessarily supplemented by the eight chromatic $W$~bosons, $W^{\tc}~|~{\tc\in\{1,\ldots,8\}}$. Off-diagonal, it is often also convenient to work with the %
representations $e_{ij}^C~|~i\not=j$, \draftone{which are associated with the elementary matrices $e_{ij}$. For example,}
$e^C_{12}=\frac{1}{\sqrt{2}}(\lambda^C_1+\rmi\lambda^C_2)$. 
\draftone{In keeping with operator notation, \prm{W^{\dot cc}} is used to denote a negatively-charged complex chromatic \prm{W} boson while \prm{{W^{\dot cc}}^\dagger} denotes a positively-charged complex chromatic \prm{W} boson. It is usually clear from context whether a notation such as \prm{W} or \prm{W^{\dot cc}} refers to both the positively- and negatively-charged species collectively or whether it specifically represents the negatively-charged species. %
The explicit charge notation \prm{W^-} and \prm{W^+} (equivalently \prm{W^{\dot{c}c -}} and \prm{W^{\dot{c}c +}}) is used as needed for clarity.}

Chromatic $Z$ bosons may likewise be introduced, as may chromatic photons though the chromatic photons remain massless \schapnotchap{sec:colouredAmass}{pfeifer2022CASM4}.

\subsubsection{Mass mechanisms in CASMIR\label{sec:CASMIRmass}}

To understand the circumstances in which chromatic $W$~bosons are---and are not---encountered in CASMIR, it is first necessary to understand the mass mechanism of CASMIR and how it differs from the Standard Model.

In the Standard Model, mass is hypothesized to arise from interaction with a vacuum state with nonvanishing Higgs expectation value. A suitable scalar boson has been observed~\cite{chatrchyan2012,aad2012}, lending support to this construction. In contrast, mass in CASMIR arises from nonvanishing expectation values of %
products of the fermion, photon, and gluon fields of the pseudovacuum as well as the scalar boson field, with the latter making only a small contribution. %
Nevertheless, %
calculations performed in CASMIR are generally quite capable of reproducing Standard Model and experimental results, and to this end, CASMIR may be considered a useful analogue to the Standard Model.

Considering specifically the achromatic and chromatic $W$~bosons, the achromatic $W$~boson acquires mass from the fermion, photon, and scalar sectors of the pseudovacuum only, while the chromatic $W$~boson species also couple to the gluon sector of the pseudovacuum. The chromatic $W$~bosons thus acquire a slightly higher mass than their achromatic counterparts, though the contribution which can arise from gluon couplings in CASMIR is known to be small compared with that from the fermion sector~\cchapnotchap{ch:detail}{pfeifer2022CASM4}. 

The energy scale of the pseudovacuum excitations which engage in these mass-generating couplings is 
\begin{equation}
\mc{E}_0=3.5864883(17)~\GeV, \label{eq:E0Ch8} %
\end{equation}
a parameter which is uniquely determined from the fine structure constant and the masses of the electron and muon \cchapnotchap{ch:detail}{pfeifer2022CASM4}. The associated timescale, being the timescale over which the pseudovacuum appears on-average homogeneous in its isotropy frame,
is denoted $t_0$:%
\begin{align}
\begin{split}
t_0&=\frac{h}{|e|}\cdot[3.5864883(17)~\GeV]^{-1}\\&\approx 1.2\times 10^{-24}~\mrm{s}. %
\end{split}\label{eq:t0}
\end{align}
Excitations in the pseudovacuum overlap,%
\footnote{More precisely, excitations in the CASMIR pseudovacuum correspond to gradients in the product of a collection of underlying (``fundamental'') scalar fields. These fields are myriad, but local behavior is dominated by those excitations with inflection points within a local region characterized by scale \prm{\mc{L}_0=ct_0}. These regions of primary influence overlap, with an average of \prm{N_0} such fields exerting significant influence at any given point in space--time. Evaluation of the mass interactions of CASMIR yields an optimal value of \prm{N_0\approx 192} for emulation of the Standard Model\notchap{.\pcite{pfeifer2022CASM4}}\chap{~(\aref{apdx:accessory}).} See also \pAref{apdx:massrel} for the equations which fix the value of \prm{N_0}.} and the mean time between pseudovacuum scattering events is denoted $t_\Omega$%
, associated with an energy scale $\mc{E}_\Omega$,
\begin{align}
\mc{E}_\Omega&=6.17960(12)~\TeV\label{eq:EOmega}\\ %
\begin{split}
t_\Omega&= \frac{h}{|e|}\cdot[6.17960(12)~\TeV]^{-1}\\&\approx 6.7\times 10^{-28}~\mrm{s}. %
\end{split}
\label{eq:tOmega}
\end{align}
Construction of an effective massive propagator requires sequential couplings to the pseudovacuum over multiple intervals $t_\Omega$ in the pseudovacuum isotropy frame. 

\subsubsection{Environments supporting chromatic bosons in CASMIR\label{sec:chromenv}}

With the mass of the chromatic $W$~boson anticipated to be only a little higher than that of the achromatic $W$~boson, \draftone{access to} both chromatic and achromatic $W$~\draftone{boson mass eigenstates} may be anticipated in high-energy environments such as the $1.96~\TeV$ collisions of the CDF~II experiment. However, the chromatic $W$~boson is a colored particle whereas the total color of a $p\bar p$ collision vanishes. In an environment containing few color charges a chromatic $W$~boson is tightly bound to its color-neutralising counterparts, and gluon exchange distributes energy and momentum among these colored species over the preon confinement timescale $t_\preon$. These few-particle confinement interactions do not in general cancel one another out, and if they occur on a similar or shorter timeframe than pseudovacuum scattering events, they disrupt the construction of massive propagators and cause confined colored bosons to behave as if massless.

In contrast, in a sufficiently homogeneous and anisotropic color-rich environment with vanishing net particle color due to a balanced sum over color charges, confinement is weakened and a colored boson acquires asymptotic freedom. Provided the extent of the environment is large compared with $t_\preon$, a chromatic $W$~boson may propagate freely through the colored environment while engaging in multiple sequential mass-generating interactions with the CASMIR pseudovacuum. Thus the chromatic $W$~bosons acquire mass only in color-rich environments with lifetimes long compared with $t_\preon$.
Regarding $t_\preon$ and the associated energy scale $\mc{E}_\preon$: 
\begin{itemize}
\item Since the quark color charges are residuals of the preon color charges, it follows that $\mc{E}_\preon$ is no smaller than the strong force scale in CASMIR.
\item In CASMIR, quasiparticles arise as perturbations on a collection of scalar fields whose inflection points have a characteristic space--time separation~\prm{t_\Omega} in the CASMIR isotropy frame. As a consequence the energy scale of spontaneous particle emissions is bounded from above by \prm{\mc{E}_\Omega} in this frame, and the mean time between spontaneous emission events is bounded from below by $t_\Omega$.
It thus follows that $\mc{E}_\preon$ in this frame is no larger than $\mc{E}_\Omega$.
\item There exists no mechanism in the Standard Model or CASMIR to fix the strong scale below $\mc{E}_\Omega$, nor experimental evidence to suggest this. To the extent that CASMIR reproduces the behaviors of the Standard Model (or in case of discrepancy, reproduces experimentally observed behaviors) below $\mc{E}_\Omega$, it may be assumed that the preon binding scale in CASMIR corresponds to the upper energy scale at which spontaneous gluon exchange may take place, yielding
\begin{equation}
\mc{E}_\preon=\mc{E}_\Omega\qquad t_\preon=t_\Omega=h\cdot\mc{E}_\Omega^{-1}.
\end{equation}
\end{itemize}
Color-rich environments with lifetimes long compared with $t_\preon$ are thus favored by energy scales small compared with $\mc{E}_\Omega\approx 6.18~\TeV$. %

Next recall that all fermion species in CASMIR are composed from three colored preons, and all %
bosons %
are constructed from co-propagating preon/antipreon pairs\chap{ (\cref{ch:SM})}.\notchap{\cite{pfeifer2022CASM3}} Binding in the fermions is mediated by preon exchange, in particular the exchange of preon/antipreon pairs with nonvanishing color but vanishing electroweak charge, which form a representation of $\SU{3}_C$ and act as effective gluons.

\draftone{Keeping the above in mind, it is now necessary to identify those environments which may sustain virtual chromatic weak vector bosons, and those which may sustain real chromatic weak vector bosons. To begin with virtual bosons, recall that i}n CASMIR, virtual particles borrow %
energy from the pseudovacuum. The typical energy imparted by a single pseudovacuum interaction is on order $\mc{E}_0$, but $\ILO{N_0}$ %
consecutive interactions within the pseudovacuum autocorrelation time $t_0$ may readily result in net foreground energy fluctuations with magnitudes in the range $[0,\ILO{\mc{E}_\Omega}]$ (with rapidly diminishing but nonvanishing probability of finding fluctuations above this scale).
Over length or time scales small compared with the pseudovacuum autocorrelation scale $t_0$ but large compared with the preon binding scale $t_\Omega$, the environment of these composite particles is naturally color-rich due to the intrinsic inhomogeneity of the pseudovacuum over these scales, effectively manifesting as a large number of extant virtual particles arising from pseudovacuum interactions. However, over timescales larger than $t_0$ the large-scale homogeneity of the pseudovacuum dominates, and the environment appears net colorless with localized fluctuations in color occuring only over timescales within the range $[t_0,t_\Omega]$. Thus the pseudovacuum environment %
only appears color-rich to species at energy scales~$\mc{E}$ satisfying
\begin{equation}
3.59~\GeV\approx\mc{E}_0<\mc{E}<\mc{E}_\Omega\approx 6.18~\TeV.\label{eq:Ebounds}
\end{equation}

It must, however, be noted that %
$\mc{E}$ is not in general the centre-of-mass energy of the collider. First, recognize that the energy scale of individual collisions is much smaller than the collider energy scale, with useful $W$~boson-forming events typically occurring at sub-$\TeV$ collision energies in both ATLAS and CDF~II~\cite{aaboud2018,aaltonen2022}. %
Second, recognize that energy scale $\mc{E}_0$ corresponds to the inverse of the threshold scale for autocorrelation of the pseudovacuum, denoted by $\mc{L}_0$ in \notchap{Refs.~\citen{pfeifer2022CASM1}, \citen{pfeifer2022CASM4}, \citen{pfeifer2022CASM2}, and \citen{pfeifer2022CASM3},}\chap{\crefr{ch:simplest}{ch:detail}} and by $t_0$ in \sref{sec:CASMIRmass} above, and represents the smallest distance or time which can be consistently discriminated from zero by measurements performed on the pseudovacuum state. For a \draftone{particle} %
energy $E$ in the isotropy frame of the pseudovacuum, the smallest time over which measurements on the accelerated particles are similarly consistently correlated (and not anticorrelated) is given by the half-period %
\begin{equation}
t_E=\frac{h}{2E}. \label{eq:tE}
\end{equation}
The energy scale $\mc{E}$ is therefore best identified with $2E$, and the energy must satisfy
\begin{equation}
3.59~\GeV\approx\mc{E}_0<2E<\mc{E}_\Omega\approx 6.18~\TeV,\label{eq:Ebounds2}
\end{equation}
i.e.~
\begin{equation}
1.79~\GeV%
<E<%
3.09~\TeV.\label{eq:Ebounds3}
\end{equation}
\chap{For more detail see \Aref{apdx:factwo}.}
As a convention, foreground field energy scales are %
denoted by $E$ to distinguish them from pseudovacuum energy scales denoted by $\mc{E}$.

\draftone{The above criterion is sufficient for formation of virtual chromatic weak vector bosons, and is necessary where a color dipole is spontaneously generated (i.e.~as a result of color-inhomogeneous pseudovacuum interactions). There then exists a second criterion which is %
necessary for the formation of real chromatic weak vector bosons: %
the existence of a real color-rich environment to which the chromatic boson is exposed for a duration sufficiently long compared with the minimum pseudovacuum interaction timescale $t_\Omega$. In practice a duration of $2t_\Omega$ suffices, %
such that for any mass vertex occurring at a time~$t$ within an interval $I=[0,2t_\Omega)$ there exists a second correlated vertex, nominally at a separation of~$\pm t_\Omega$, which also lies within interval~$I$.}

\draftone{The simplest environment providing an appropriate color-rich foreground environment is the interior of a hadron. With the $W$~bosons in CDF~II and ATLAS being generated close to rest in the centre-of-mass frame, their durations of exposure to this color-rich environment are determined by the energies of the protons or antiprotons in this frame,
\begin{equation}
E_p=\frac{\sqrt{s}}{2}.
\end{equation}
The requirement that the color-rich environment endures for a duration of $2t_\Omega$ rather than $t_\Omega$ tightens the upper bound of \Eref{eq:Ebounds3} to yield %
\begin{equation}
\draftone{3.59~\GeV\approx \mc{E}_0<2E_p< \frac{\mc{E}_\Omega}{2}\approx 3.09~\TeV},\label{eq:fgconstraint_pre}
\end{equation}
or equivalently, %
\begin{equation}
\draftone{3.59~\GeV\approx \mc{E}_0<\sqrt{s}< \frac{\mc{E}_\Omega}{2}\approx3.09~\TeV}.\label{eq:fgconstraint}
\end{equation}
The protons at Tevatron satisfy \Erefs{eq:fgconstraint_pre}{eq:fgconstraint}, having energies $E_p=0.98~\TeV$ and $\sqrt{s}=1.96~\TeV$. In contrast the $W$~mass dataset at ATLAS %
was collected at $\sqrt{s}=7~\TeV$ ($E_p=3.5~\TeV$) and does not satisfy these constraints. 
The chromatic weak vector boson mass eigenstates are therefore %
accessible to CDF~II but not to ATLAS.}

\draftone{%
As a further corollary of these %
considerations, the chromatic weak vector boson mass eigenstates are also inaccessible to lepton colliders such as LEP as their collision environments lack a %
color-rich (hadronic) foreground.}

\subsection{The chromatic $W$ boson\label{sec:chrWbos}} %

\subsubsection{In CDF~II\label{sec:inCDFII}}

Measurement of the mass of $W$~bosons produced in proton--proton or proton--antiproton collisions relies on the detection of charged leptons produced
during decay processes such as
\begin{equation}
W\longrightarrow e_L+\bar\nu_e
\end{equation}
and higher-generation counterparts. Given a distribution of energies and momenta on the emitted electron, 
the rest mass of the (achromatic) $W$~boson may then be reconstructed. 
In CASMIR, however, left-helicity electrons may \draftone{in theory} also be generated by 
the decay of chromatic $W$~bosons.
As noted in \sref{sec:chromenv}, the energy scale of these production events in CDF~II favors interpretation of the pseudovacuum as a color-rich environment, and thus when modeled in CASMIR it is well-suited to the creation of chromatic as well as achromatic $W$~bosons.

In terms of the parameters of the CASMIR mass relationships (see \Aref{apdx:massrel} or \notchap{\rcite{pfeifer2022CASM4}, }\pcref{ch:detail}), the squares of the masses of the achromatic massive bosons are proportional to factors which may be written in the form %
\begin{equation}
\left.1+\frac{z}{18\left[k^{(e)}_{1}(\mc{E}_e)\,{N_0}\right]^4}\qquad \right|%
\qquad z\in\mbb{Z}^+
\end{equation}
where $z=19$ for $W$~bosons and $z=55$ for $Z$~bosons. When colored $W$ and $Z$ bosons couple to the gluon component of the pseudovacuum this increases the numerator $z$ to $155\frac{11}{16}$ and $\draftone{62\frac{11}{64}}$ respectively.
The ratio of masses of the chromatic and achromatic $W$~bosons is consequently
\begin{equation}
\left(\frac{m_{W^{\tc}}}{m_W}\right)^2=\frac
{1+\frac{2491}{288\left[k^{(e)}_{1}(\mc{E}_e)\,{N_0}\right]^4}}
{1+\frac{19}{18\left[k^{(e)}_{1}(\mc{E}_e)\,{N_0}\right]^4}}
\left(1+\mc{O}_b\right)\draftone{.}
\end{equation}
\draftone{In conjunction with the equations of \Aref{apdx:massrel}, this} yields achromatic and chromatic boson masses
\begin{align}
m_W &= 80.3587(22)~\GeV/c^2\label{eq:Wmass}\\ 
m_{W^{\tc}\draftone{|_{\tc<9}}} &= 80.4434(22)~\GeV/c^2 %
\label{eq:Wtcmass}
\end{align}
\draftone{where $m_W$ is the achromatic $W$ boson mass associated with the trivial $C$-sector representation $\lambda^C_9$.}
\draftone{The other eight chromatic $W$ boson mass eigenvalues $1\leq\tc\leq 8$ corresponding to the 8-dimensional representation of $\su{3}_C$ are degenerate, as the $\SU{3}_C$ symmetry of the color sector is unbroken by construction\notchap{~\cite{pfeifer2022CASM3}.}\chap{ (see \cref{ch:SM}).}}

\begin{figure}
\includegraphics[width=\linewidth]{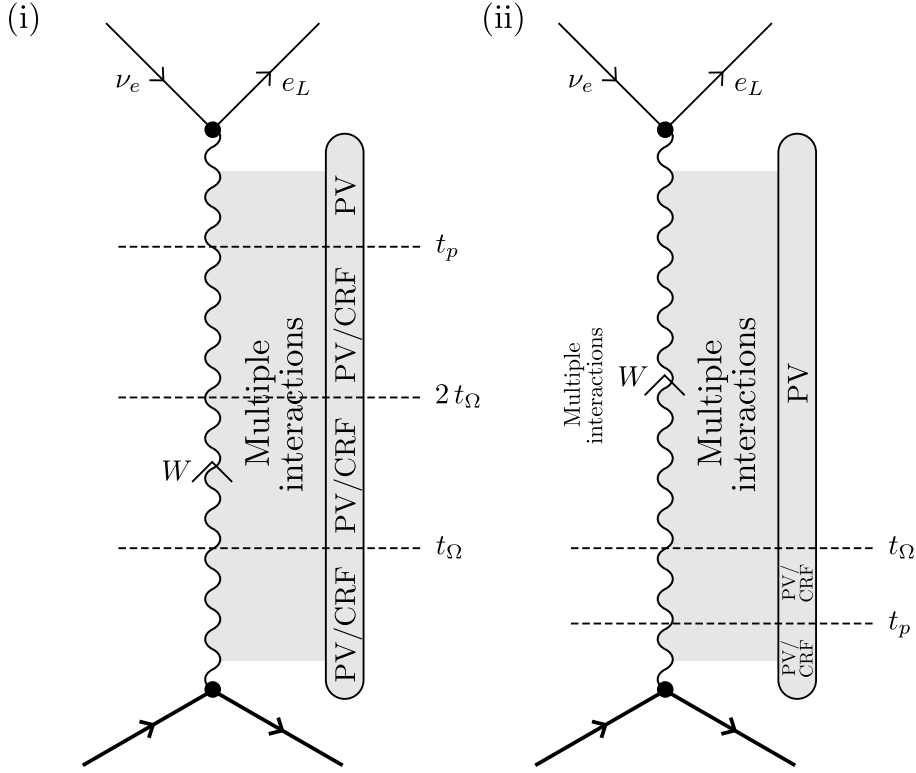}
\caption{\draftone{(i)~Heuristic representation of pseudovacuum~(PV) and color-rich foreground~(CRF) interactions of a weak vector boson having \prm{W}-type character on the \prm{A}~sector in CDF~II. (ii)~Heuristic representation of PV and CRF interactions of a weak vector boson having \prm{W}-type character on the \prm{A}~sector in ATLAS. Time \prm{2t_\Omega}~\peref{eq:tOmega} is the minimum length scale reliably supporting a mass-squared interaction, and \prm{t_p} is the timescale associated with the energy \prm{\frac{1}{2}\sqrt{s}} of each proton or antiproton beam~\eref{eq:tp}. In any hadron collision, the confinement scale---and hence the extent of the CRF---is determined the greater of \prm{t_\Omega} and \prm{t_p} in the pseudovacuum isotropy frame, which approximates the centre of mass frame. The grey shaded area represents multiple particle exchanges of unspecified type between the propagating boson, the pseudovacuum, and (where present) the color-rich foreground.}\label{fig:Wprop}}
\end{figure}

\draftone{In practice %
the chromatic $W$ boson mass eigenstates $1\leq\tc\leq 8$ %
are not observed directly in CDF~II. 
When a $p\bar{p}$ collision generates a particle carrying $A$-sector representation $e^A_{23}$ (corresponding to some version of the $W$~boson) or its conjugate, this propagates within the color-rich environment of the collision %
for a duration $t_p$ which is characterized by the per-hadron energy $\frac{\sqrt{s}}{2}$. 
In this context
a factor of $\frac{1}{2}$ arises because $W$~boson formation occurs on average half-way through the time period for which the site of formation lies within the color-rich environment. 
The expression for the mean time that the $W$~boson exists within a color-rich environment is therefore
given by
\begin{equation}
t_p=\frac{h}{2}\cdot\frac{2}{\sqrt{s}}=\frac{h}{|e|}\cdot [1.96~\TeV]^{-1} %
\label{eq:tp}
\end{equation}
which satisfies
\begin{equation}
t_p> 2t_\Omega. \label{eq:tpconstraint}
\end{equation}
During period~$t_p$ the boson %
undergoes multiple mass-generating interactions with the pseudovacuum and mutiple color-exchange interactions with the foreground environment as illustrated in \fref{fig:Wprop}(i).}
\draftone{%
While individual mass vertices may be characterized as chromatic or achromatic, recall that at the preon scale all scattering interactions involve exchange of particles carrying both $A$-sector and $C$-sector charges, and in a color-rich foreground environment a boson associated with representation $\lambda^C_9$ may undergo $A$-sector-mediated interactions which as a side-effect change its representation on the $C$~sector.}
\draftone{An individual boson may therefore be considered to experience chromatic mass vertex interactions and achromatic mass vertex interactions $\frac{8}{9}$ and $\frac{1}{9}$ of the time respectively, being associated with factors of $m_{W^{\tc}}^2$ and $m_W^{2}$. For a $W$-type boson in a color-rich environment the mean mass-squared per vertex over period $t_p$ is therefore given by
\begin{equation}
\left([m_W]_\mrm{r.m.s.}\right)^2:=\frac{1}{9}\sum_{\tc\in\{1,\ldots,9\}} m_{W^{\tc}}^2
\end{equation}
where $m_{W^9}$ corresponds to the achromatic $W$ boson mass $m_W$.} %
\draftone{The mass $[m_W]_\mrm{r.m.s.}$} equates to $m_{W^{\dot cc}}$ in \sref{sec:colouredWmass}\notchap{ of \rcite{pfeifer2022CASM4}}\draftone{, and substituting for $m_W$ and $m_{W^\tc}$%
} yields
\begin{equation}
[m_W]_\mrm{r.m.s.}=80.4340(22)~\GeV/c^2.
\end{equation}
This expression receives a small energy-dependent correction due to the difference in mass of the different $W$~species, which 
is assumed %
to be negligible at the energy scales and precision of the current \paper{}.

\draftone{Since $t_p$ in CDF~II is short compared with the lifetime of the $W$~boson [total width $2.085(42)~\GeV$]~\cite{workman2022}, there is then a transition at the end of time period $t_p$ from a color-rich to a color-poor environment. Prior to this transition the $W$-type boson undergoes multiple foreground interactions in which it may gain or lose momentum, and its preferred mean rest energy is determined by the mass vertices it has experienced up to this point, corresponding to $[m_W]_\mrm{r.m.s.}$. For the boson to escape confinement, it must be color-neutral on transitioning from $t< t_p$ to $t\gg t_p$, a criterion which implies the recapture of $\frac{8}{9}$ of $W$-type bosons. Although those which escape are now off-shell, being now subject only to achromatic ($\lambda^C_9$) mass vertices, they remain in an excited state because
they no longer exist in a context of multiple foreground scattering interactions, and so are limited in their ability to shed this $75~\MeV$ %
of surplus energy. Candidate processes such as $W^*\rightarrow W\gamma\gamma$ are characterized by the timescale of the surplus, and thus are slow when compared with the usual $W$~boson decay modes. The excited $W$-type bosons therefore decay via the usual channels, including the dilepton channels
\begin{align}
\begin{split}
(W^{-})^*&\longrightarrow e_L+\bar\nu_e\\
(W^{+})^*&\longrightarrow \bar{e}_L+\nu_e.
\end{split}
\end{align} 
As usual the (anti)neutrino carries away missing transverse momentum, while the charged lepton energy distribution reflects the off-shell, excited mass of 
\begin{equation}
[m_W]_\mrm{CDF\,II}=[m_W]_\mrm{r.m.s.}=80.4340(22)~\GeV/c^2.
\end{equation}
}

\draftone{Note that the recapture of $\frac{8}{9}$ of $W$-type bosons cancels the ninefold increase in number of $W$-boson-generating channels, and hence the predicted rate of $W$~boson production is not significantly changed from that of the Standard Model \drafttwo{(see also \sref{sec:sxBR})}.}

\subsubsection{In ATLAS\label{sec:WcinATLAS}}

Now compare %
with the measurement of $W$~\draftone{boson} mass in the ATLAS experiment~\cite{aaboud2018,aaboud2018a}. Collection of the existing ATLAS dataset took place at a collider energy of $E_\mrm{ATLAS}\,\draftone{:\!}=\draftone{\sqrt{s}=}~7~\TeV$, again with individual $W$~boson formation events being \mbox{sub-$\TeV$}%
.
The energy of $W$~boson formation in ATLAS therefore falls within the interval in which the pseudovacuum appears color-rich,
\begin{equation}
3.59~\GeV\approx\mc{E}_0<2E<\mc{E}_\Omega\approx 6.18~\TeV,\tagref{eq:Ebounds2}
\end{equation}
\draftone{but the centre-of-mass energy $\sqrt{s}=7~\TeV$ does not satisfy the color-rich foreground constraint
\begin{equation}
\draftone{3.59~\GeV\approx \mc{E}_0<\sqrt{s}< \frac{\mc{E}_\Omega}{2}\approx3.09~\TeV}.\tagref{eq:fgconstraint}
\end{equation}
As noted in \sref{sec:chromenv} confinement in CASMIR operates over a minimum timescale of $t_\Omega$ in the pseudovacuum isotopy frame,
so the duration of the color-rich collision environment in ATLAS is given by
\begin{equation}
\frac{h}{|e|}\min{(\sqrt{s},\mc{E}_\Omega)}^{-1}
\end{equation}
rather than $\frac{h}{|e|}\sqrt{s}$, but even on substituting $\mc{E}_\Omega$ for $\sqrt{s}$ in \Eref{eq:fgconstraint}
the lifetime of the color-rich foreground is seen to be insufficient to support chromatic weak vector boson mass vertices. The LHC at $7~\TeV$ therefore generates only achromatic massive weak vector bosons, with CASMIR predicting an observed $W$~boson mass at ATLAS of} %
\draftone{%
\begin{equation}
m_W = 80.3587(22)~\GeV/c^2\tagref{eq:Wmass}.
\end{equation}
}

\draftone{%
It is also worth noting}
a \draftone{further} key difference in the selection of $W$~boson decay events for the ATLAS calculation%
.
To reduce background contributions at energy scales of interest, 
the ATLAS evaluations of $W$~\draftone{boson} mass %
explicitly exclude events in which the charged lepton is closely spatially related to a hadron jet:\cite{the-ATLAS-collaboration2023,aaboud2018}
\begin{itemize}
\item{} ``Additional isolation requirements on the nearby activity in the ID and calorimeter are applied to improve background rejection. These isolation requirements are implemented by requiring the scalar sum of the $p_\mrm{T}$ of tracks in a cone of size $\Delta R\equiv\sqrt{(\Delta\eta)^2+(\Delta\phi)^2}<0.4$ around the electron, $p^{e,\mrm{cone}}_\mrm{T}$, and the transverse energy deposited in the calorimeter within a cone of size $\Delta R<0.2$ around the electron, $E^\mrm{cone}_\mrm{T}$, to be small.''\cite{aaboud2018}
\item{} ``Similarly to the electrons, the rejection of multijet background is increased by applying an isolation requirement~: the scalar sum of the $p_\mrm{T}$ of tracks in a cone of size $\Delta R<0.2$ around the muon candidate, $p^{\mu,\mrm{cone}}_\mrm{T}$, is required to be less than 10\% of the muon $p_\mrm{T}$.''\cite{aaboud2018}
\end{itemize}
\draftone{If color charges are entrained along the trajectory of the $W$~boson prior to it crossing the confinement scale, it is possible that this may bias the direction of jet emission to align with the hypothesized mass-enhanced $W%
$~bosons. The effect of apparent rest mass enhancement is not anticipated to depend on the direction of jet emission, and although this cut may result in exclusion of a larger-than-anticipated number of $W$~boson decay events at energies $\sqrt{s}<\frac{1}{2}\mc{E}_\Omega$, it is 
not anticipated to have a significant impact on the value measured for the apparent rest mass of the $W$~boson.
}

\begin{figure}
\centering
\includegraphics[width=0.7\linewidth]{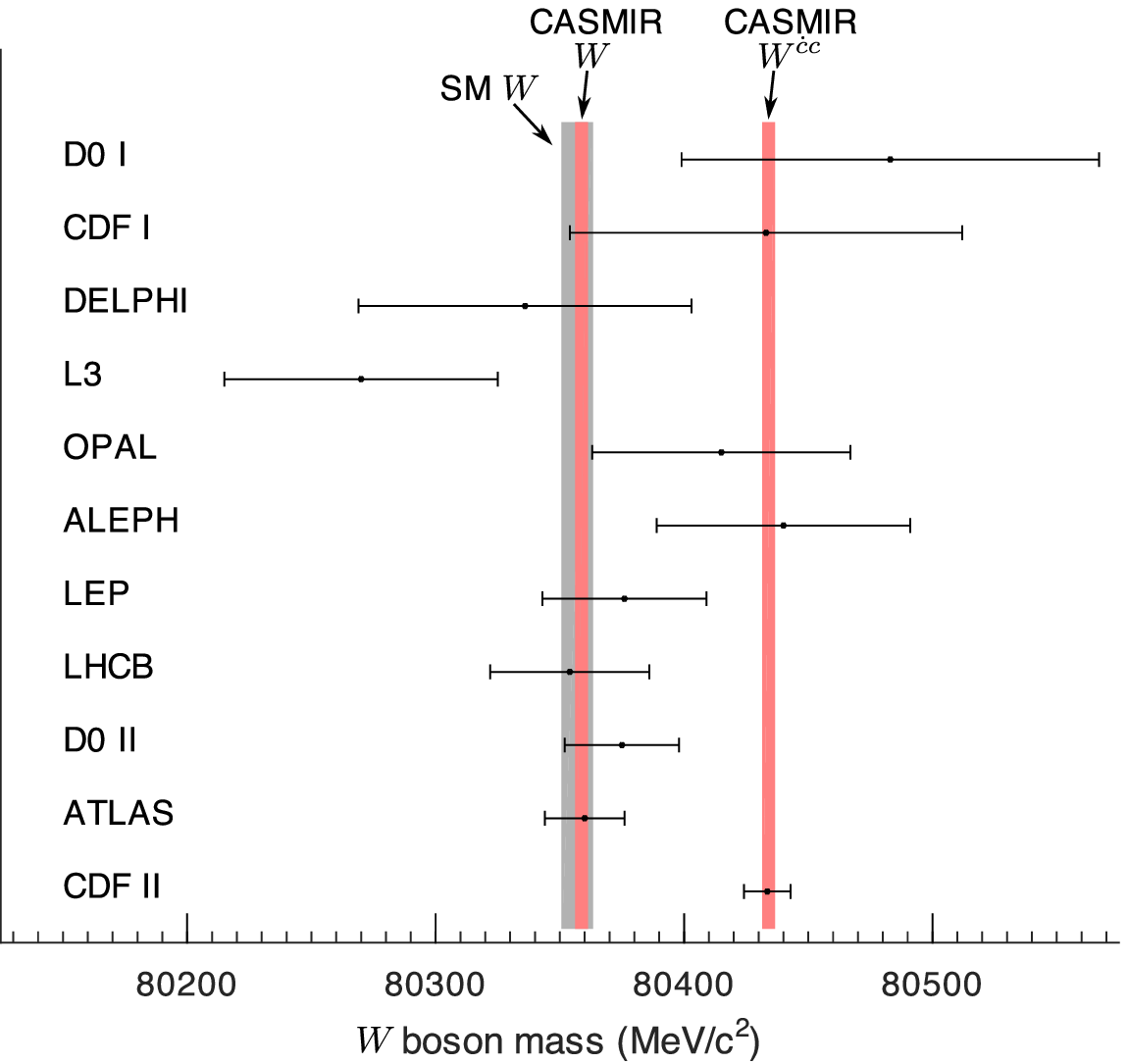}
\caption{Comparison of \prm{W}~boson mass measurements\chap{~}\pcite{aaltonen2022,the-ATLAS-collaboration2023,abazov2012,aaij2022a,schael2013,schael2006,abbiendi2006,achard2006,abdallah2008,affolder2001,abazov2002} with predictions of the Standard Model (SM~\prm{W})\notchap{,}\chap{~}\pcite{awramik2004,erler2019,workman2022}\chap{,} the CASMIR achromatic \prm{W}~boson (CASMIR~\prm{W}), and \draftone{the CASMIR mixed chromatic/achromatic \prm{W}~boson} (CASMIR~\prm{W^{\dot cc}}). ``LEP'' denotes the average value from LEP\chap{~}\notchap{.}\cite{schael2013}\chap{.} %
\label{fig:Wmass}}
\end{figure}

Measurements and predictions for $W$~boson masses are summarized in \tref{tab:Wresults}.
\begin{table}
\tbl{Comparison of predictions from CASMIR with \prm{W}~boson mass measurements from CDF~II (Tevatron)\notchap{\draftone{,}}\chap{~}\pcite{aaltonen2022}\chap{,} ATLAS (LHC)\notchap{\draftone{,}}\chap{~}\pcite{the-ATLAS-collaboration2023}\chap{,} \draftone{and LEP\chap{~}\notchap{.}\pcite{schael2013}\chap{.}}
The Standard Model (SM) prediction for the \prm{W}~boson mass is \prm{80.356(6)~\GeV/c^2}\notchap{,}\chap{~}\pcite{awramik2004,erler2019,workman2022}\chap{,} while the CASMIR predictions for the achromatic \prm{(W)} and mixed chromatic/achromatic \prm{(W^{\dot cc})}~boson masses are \prm{80.3587(22)~\GeV/c^2} and \prm{80.4340(22)~\GeV/c^2} respectively.  As noted in \Aref{apdx:massrel}, mass uncertainties in CASMIR are likely overestimated. To avoid unfairly advantaging the CASMIR results, an upper bound on tensions is also obtained by taking the CASMIR uncertainty to zero; where this upper bound differs to the calculated tension, it is reported after the tension in brackets. Tensions smaller than \prm{1\,\sigma} are displayed in green online.}
{
\chap{\begin{center}}\chapeight{\begin{center}}
\begin{tabular}{@{}llllll@{}}
\toprule
Experiment&Predicted&Mass observed&Tension with&Tension with&Tension with\\
&$W$ boson \draftone{mass}&\prm{(\GeV/c^2)}&SM $W$&CASMIR $W$&CASMIR $W^{\dot cc}$\\
\midrule
CDF~II & mixed ($W^{\dot cc}$) & $80.4335(94)$ & $6.9\,\sigma$ & $7.7\,\sigma~(\leq 8.0\,\sigma)$ & \textcolor{darkgreen}{$0.1\,\sigma$}\\
ATLAS & achromatic ($W$) &$80.360(16)$ & \textcolor{darkgreen}{$0.2\,\sigma$} & \textcolor{darkgreen}{$0.1\,\sigma$} & $4.6\,\sigma$ \\
\draftone{LEP} & \draftone{achromatic ($W$)} &\draftone{$80.376(33)$} & \textcolor{darkgreen}{$0.6\,\sigma$} & \textcolor{darkgreen}{$0.5\,\sigma$} & \draftone{$1.8\,\sigma$}\\
\botrule
\end{tabular}\label{tab:Wresults}
\chap{\end{center}}\chapeight{\end{center}}
}
\end{table}
Further comparisons with other measurements of $W$~boson mass %
are shown in \fref{fig:Wmass}. 
If these results %
are compared with the predictions of CASMIR then with the exception of L3 at $1.6\,\sigma$, all results have tension $0.7\,\sigma$ or less with the CASMIR predicted mass for either the achromatic or mixed chromatic/achromatic $W$~boson species. %

\subsubsection{In LEP}

It is next appropriate to ask whether CASMIR might also predict observation of a mixed chromatic and achromatic $W$ boson signal in the combined pair production data from LEP (see ALEPH, DELPHI, L3, OPAL, and the combined LEP value in \fref{fig:Wmass})~\cite{abbiendi2006,schael2006,abdallah2008,achard2006,schael2013}. 
The widths of the LEP data error bars prohibit clean discrimination between chromatic and achromatic boson signals, such that even for the combined LEP dataset, which is most consistent with the achromatic $W$ boson mass $m_W$, tension with the mixed chromatic and achromatic value $m_{W^{\dot cc}}$ is just $1.8\,\sigma$, and is dominated by experimental uncertainty.
The LEP datasets therefore make no assessment as to whether CASMIR provides a qualitatively good effective description of observed weak boson mass. %
\draftone{Recognising that lepton collisions in LEP do not occur within a spatially extensive color-rich environment as per \sref{sec:chromenv}, CASMIR predicts achromatic massive weak vector boson formation in lepton colliders such as LEP.}

\subsection{The chromatic $Z$ boson\label{sec:chrZbos}}

There is also a further %
possibility for testing the predictive power of CASMIR in the weak sector.
Similar to the chromatic $W$~bosons described above, CASMIR also necessarily incorporates the existence of chromatic $Z$~bosons in color-rich environments at energies below $\mc{E}_\Omega$. 
\draftone{Repeating the boson mass calculations of \sref{sec:chrWbos} for $Z$~bosons while solving the mass relationships of \Aref{apdx:massrel} yields 
\begin{align}
\left(\frac{m_{Z^{\tc}}}{m_Z}\right)^2&=\frac
{1+\frac{3979}{1152\left[k^{(e)}_{1}(\mc{E}_e)\,{N_0}\right]^4}}
{1+\frac{55}{18\left[k^{(e)}_{1}(\mc{E}_e)\,{N_0}\right]^4}}
\left(1+\mc{O}_b\right)\\
\left([m_Z]_\mrm{r.m.s.}\right)^2&=\frac{1}{9}\sum_{\tc\in\{1,\ldots,9\}} m_{Z^{\tc}}^2\\
m_{Z^\tc|_{\tc<9}}&=91.1928(35)~\GeV/c^2\\
m_Z&=91.1877(35)~\GeV/c^2\\ 
[m_Z]_\mrm{CDF\,II} &= [m_Z]_\mrm{r.m.s.} = 91.1922(35)~\GeV/c^2.\label{eq:mZrms}
\end{align}
As was seen for chromatic $W$ bosons of \sref{sec:chrWbos}, a chromatic $Z$ boson is not independently detectable using existing apparatus, but the existence of the chromatic $Z$ boson mass vertices within a sufficiently long-lived color-rich collision environment causes achromatic $Z$~bosons to be emitted in an excited off-shell state. In contrast with $W$~boson emission, when a $Z$~boson decays via a charged dilepton channel both decay products may be recovered, permitting higher precision measurements of boson mass to be achieved. However, the separation of the normal and excited energy levels is more than an order of magnitude smaller than for the $W$~boson, at only $4.5~\MeV$. Comparison of theory and experiment for CDF~II and LEP is presented in \tref{tab:Zresults}, but tensions are small and no significant conclusions can be drawn.
}
\begin{table}
\tbl{\draftone{Comparison of predictions from CASMIR with \prm{Z}~boson mass measurements from CDF~II (Tevatron)%
\chap{~}\pcite{aaltonen2022}%
 and LEP\chap{~}\notchap{.}\pcite{schael2013}\chap{.}
CASMIR predictions for the achromatic \prm{(Z)} and mixed chromatic/achromatic \prm{(Z^{\dot cc})}~boson masses are \prm{91.1877(35)~\GeV/c^2} and \prm{91.1922(35)~\GeV/c^2} respectively.  As noted in \Aref{apdx:massrel}, mass uncertainties in CASMIR are likely overestimated. To avoid unfairly advantaging the CASMIR results, an upper bound on tensions is also obtained by taking the CASMIR uncertainty to zero; where this upper bound differs to the calculated tension, it is reported after the tension in brackets. Tensions smaller than \prm{1\,\sigma} are displayed in green online.}}
{
\chap{\begin{center}}\chapeight{\begin{center}}
\begin{tabular}{@{}lllll@{}}
\toprule
\draftone{Experiment}&\draftone{Predicted}&\draftone{Mass observed}&\draftone{Tension with}&\draftone{Tension with}\\
&\draftone{$Z$ boson mass}&\draftone{\prm{(\GeV/c^2)}}&\draftone{CASMIR $Z$}&\draftone{CASMIR $Z^{\dot cc}$}\\
\midrule
\draftone{CDF~II} & \draftone{mixed ($Z^{\dot cc}$)} & \draftone{$91.1920(75)$} & \textcolor{darkgreen}{$\p{<}\;0.5\,\sigma~(\leq 0.6\,\sigma)$} & \textcolor{darkgreen}{$<0.1\,\sigma$}\\
\draftone{LEP} & \draftone{achromatic ($Z$)} & \draftone{$91.1876(21)$} & \textcolor{darkgreen}{$<0.1\,\sigma~(\leq 0.1\,\sigma)$} & \draftone{$\p{<}\;1.1\,\sigma~(\leq 2.2\,\sigma)$}\\
\botrule
\end{tabular}\label{tab:Zresults}
\chap{\end{center}}\chapeight{\end{center}}
}
\end{table}

Fascinatingly, in addition to data collected by CDF~II there also exists one other dataset which is already sensitive to signals consistent with chromatic $Z$~bosons (as well as chromatic $W$~bosons), and which returns a result consistent with their existence. 
This is the Muon~\mbox{$g-2$} experiment at Fermilab~\cite{aguillard2023a}, which is discussed \drafttwo{in \sref{sec:muong2}}. %

\subsection{\drafttwo{Cross-sections and branching ratios}\label{sec:sxBR}}

\drafttwo{In addition to having effects on the measured values of $W$~and $Z$ boson mass, the differences between the electroweak sectors of CASMIR and the Standard Model also affect the massive weak vector boson cross-sections and branching ratios. This Section evaluates the magnitudes of these effects.} %

\subsubsection{\drafttwo{CASMIR corrections to Standard Model calculations}}

\drafttwo{In hadron colliders such as Tevatron and the LHC, the primary mode of $W$~and $Z$ boson production is described by the Drell-Yan process~\cite{drell1970,drell1970a} shown at tree level in \fref{fig:drellyan}.
\begin{figure}
\includegraphics[width=\linewidth]{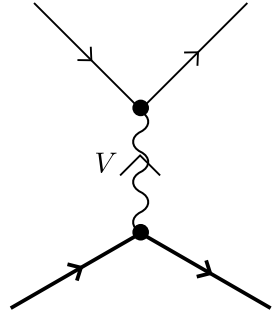}
\caption{\drafttwo{Drell-Yan process for formation of a massive weak vector boson \prm{V\in\{W^\pm,Z\}} from the collision of two partons (lower vertex) and decay into a fermion pair (upper vertex).}\label{fig:drellyan}}
\end{figure}%
Higher-order corrections to this process are well-understood, with the most important perturbative corrections arising from Quantum Chromodynamics (QCD). These have been evaluated to next-to-next-to-next-to-leading order (N$^3$LO) in \rcitet{duhr2020,duhr2020a,duhr2022}, with electroweak~(EW) corrections and mixed QCD-EW corrections being evaluated in \rcitet{baur1998,baur2022,delto2020,bonciani2020}.
As a result the Standard Model values of the associated cross-sections are known to a precision of around 2\%, with uncertainty dominated by the Parton Distribution Function (PDF). %
Similarly, measurements of these cross-sections %
in ATLAS and CMS are also capable of reaching precisions of a few parts in $10^2$~\cite{aad2016,the-CMS-collaboration2015,the-CMS-collaboration2016}, permitting testing of these predictions in the LHC.} %

\drafttwo{Theoretical evaluation of these parameters is quite involved and is generally automated through use of specialist software\notchap{, e.g.~\rcite{baglio2022}}\chap{~\cite[e.g.][]{baglio2022}}. %
Fortuitously, on comparing CASMIR with the Standard Model the corrections to cross-sections and branching ratios are not large, and %
to determine the equivalent parameters in CASMIR it suffices to consider %
corrections to the tree-level process.} %

\drafttwo{The relevant differences between the electroweak sectors of CASMIR and the Standard Model are as follows:
\begin{enumerate}
\item In CASMIR there exist eight additional $W$-like and eight additional $Z$-like bosons, carrying color as well as electroweak charges.\label{sxBRitemNum}
\item The strength of the weak interaction is a derived quantity in CASMIR, differing slightly from its measured value.\label{sxBRitemGF}
\item The weak boson masses and tau mass in CASMIR are derived quantities also, differing slightly from their measured values.\label{sxBRitemMass}
\end{enumerate}
However, at tree level the Fermi constant may be written
\begin{equation}
G_F=\frac{\pi\alpha}{\sqrt{2}m_W^2\left(1-\tfrac{m_W^2}{m_Z^2}\right)}\left[1+\OO{\frac{\alpha}{\pi}}\right],
\label{eq:GF}
\end{equation}
and since the fine structure constant $\alpha$ is an input parameter for both CASMIR and the Standard Model, it follows that the corrections arising from item~\ref{sxBRitemGF} above may be considered collectively with those arising from item~\ref{sxBRitemMass}.}

\subsubsection{\drafttwo{Consequences for cross sections and decay rates}}

\drafttwo{Consider now the tree-level contribution to the inclusive scattering cross-section $\sigma^\mrm{tot}_V$ for production of a massive vector boson $V\in\{W^\pm,Z\}$, associated with the lower vertex of \fref{fig:drellyan}.
The value of $\sigma^\mrm{tot}_V$ is affected by (i)~changes in the number of boson species having a weak sector representation consistent with boson~$V$ in the Standard Model (i.e.~inclusion of colored weak vector boson species in CASMIR), (ii)~changes to the interaction strength associated with the production of these bosons compared with production of boson~$V$ in the Standard Model, and (iii)~changes to the masses of these bosons compared with boson~$V$ in the Standard Model. It is therefore necessary to consider all three enumerated differences between the electroweak sectors of CASMIR and the Standard Model.}

\drafttwo{First, consider item~\ref{sxBRitemNum} in CDF. Recognize that as described in \sref{sec:inCDFII}, during CDF Run~II the generation of weak bosons took place within a color-rich foreground which was long-lived compared with the timescale of the CASMIR mass interaction. As a consequence, in CASMIR:
\begin{itemize}
\item Parton/parton interactions are capable of yielding any of the nine choices of $C$-sector representation for boson~$V$.
\item Preon-scale interactions may transfer color between the boson and the color-rich environment of the interior of the colliding hadrons.
\item The unbroken $\GL{3}{R}$ symmetry of the CASMIR color sector ensures that any boson has a $\frac{1}{9}$ chance of carrying the neutral color charge $\lambda_9\propto\mbb{I}_3$ when attempting to escape confinement.
\item For any such boson which escapes confinement, factorization of $\GL{3}{R}_C$ [implicitly $\GL{3}{R}_C\otimes\GL{2}{C}$] into $\GL{1}{R}_C\otimes\SU{3}_C$ [again, implicitly ${}\otimes\GL{2}{C}$] as per \aref{apdx:lemmas}\notchap{ of \rcite{pfeifer2022CASM2}}
implies that the process of creating this boson [which carries a noninteracting charge on $\GL{1}{R}_C$] is in turn necessarily factorizable into independent processes on the $A$~and $C$~sectors, with the $C$~sector being overall trivial. This factorizability holds for the emission process as a whole, even if it does not necessarily hold vertex-by-vertex within the domain of the color-rich foreground.
\item Consequently, regardless of the actual $C$-sector representation associated with a $V$-type boson at the time of its initial production, if that boson escapes confinement then the net emission vertex may be rewritten as an equivalent pure $A$-sector process, and the tree-level vertex for this process is identical to that for the colorless $V$~boson (having $C$-sector representation~$\lambda_9$), up to the effects of any differences in the mass vertex coefficients (which are addressed separately---items~\ref{sxBRitemGF}-\ref{sxBRitemMass} in the enumerated list above).
\end{itemize}
Putting this all together yields nine times as many events generating $V$-type~bosons within the hadron collision environment, all of identical weight, of which only $\frac{1}{9}$ actually result in $V$~boson emission beyond the confinement scale.}

\drafttwo{Second, the Fermi constant receives a correction, and as mentioned immediately following \Eref{eq:GF} this may be re-expressed in terms of corrections to the weak boson masses. 
Contributions arise from corrections to both the $W$~and the $Z$ boson masses, though in practice
those arising from corrections to the $Z$~boson mass are negligible.}

\drafttwo{Finally, evaluate the magnitude of corrections arising from the boson and tau masses as per item~\ref{sxBRitemMass}. For CDF~II, factors of $\left(m_V^\mrm{SM}\right)^2$ in the Standard Model calculation may be replaced by %
$\left(m_{V,\mrm{r.m.s.}}^\mrm{CASMIR}\right)^2$ resulting in a multiplicative correction which, to leading order, has the form
\begin{equation}
\left\{1+\OOO{\frac{\left(m_{V,\mrm{r.m.s.}}^\mrm{CASMIR}\right)^2-\left(m_V^\mrm{SM}\right)^2}{\left(m_V^\mrm{SM}\right)^2}}\right\}.\label{eq:massshiftfacrms}
\end{equation}
The discrepancies between the CASMIR $Z$~boson masses and their Standard Model counterparts are small [see \tref{tab:masses} of \Aref{apdx:massrel}, and \Eref{eq:mZrms}], and with $G_F$ being a derived quantity in CASMIR, corrections dependent on the $W$~mass enter all processes at tree level through \Eref{eq:GF}, even for $V=Z$. It therefore suffices to replace $V\rightarrow W$ in \Eref{eq:massshiftfacrms}.
Corrections due to differences in the tau mass are also small, and contributions to $\sigma^\mrm{tot}_V$ are of sub-leading order so may be neglected.}

\drafttwo{Next consider corrections to massive weak vector boson emission in ATLAS. Provided the centre of mass frame of the collider is sufficiently close to the isotropy frame of CASMIR, the higher energy scale reduces the lifetime of the color-rich foreground to below the timescale of the CASMIR mass interaction. This eliminates color transfer between the color-rich foreground and the weak vector boson over timescales long compared with the mass interaction scale (see \sref{sec:chromenv}).
Regarding item~\ref{sxBRitemNum}, because the color-sector representations of the bosons are fixed over mass-interaction timescales, the only bosons participating in mass interactions are those which escape confinement carrying representation $\lambda_9$. However, the resulting numerical factors of~$9$ and~$\frac{1}{9}$ remain unchanged. Regarding items~\ref{sxBRitemGF}-\ref{sxBRitemMass}, this fixing of the associated color-sector representations prevents averaging of mass vertices such that $m_{V,\mrm{r.m.s.}}^\mrm{CASMIR}$ is replaced in \Eref{eq:massshiftfacrms} by $m_{V}^\mrm{CASMIR}$:
\begin{equation}
\left\{1+\OOO{\frac{\left(m_{V}^\mrm{CASMIR}\right)^2-\left(m_V^\mrm{SM}\right)^2}{\left(m_V^\mrm{SM}\right)^2}}\right\}.\label{eq:massshiftfacplain}
\end{equation}}

\drafttwo{A similar discussion applies to the evaluation of decay constants $\Gamma(V\rightarrow\ldots)$ associated with particular choices of species at the upper vertex of \fref{fig:drellyan}.
Arguments equivalent to those presented for the cross-section similarly yield multiplicative correction factors having the forms of \Eref{eq:massshiftfacrms} for CDF~II and \Eref{eq:massshiftfacplain} for ATLAS with $V\in\{W^\pm,Z\}$. 
The very small difference in tau mass between CASMIR and the Standard Model does in principle yield a correction to the some tree-level decay constants, but this is not numerically relevant at current precisions.}

\drafttwo{To evaluate the significance of the above corrections, it is convenient to rewrite the leading-order corrections of \Erefr{eq:massshiftfacrms}{eq:massshiftfacplain} in a more general form
\begin{equation}
\left[1+\Theta_{\sigma\Gamma}(s)\Xi_W(s)\right]\label{eq:ThetaXi}
\end{equation}
where 
$\Xi_W(s)$ is a process-independent function of energy satisfying
\begin{equation}
\Xi_W(s)=\left\{\begin{array}{l}
\frac{\left(m_{W,\mrm{r.m.s.}}^\mrm{CASMIR}\right)^2-\left(m_W^\mrm{SM}\right)^2}{\left(m_W^\mrm{SM}\right)^2}~~\textrm{for $m_Wc^2\ll \sqrt{s}\ll3.09~\TeV$}\\
\frac{\left(m_{W}^\mrm{CASMIR}\right)^2-\left(m_W^\mrm{SM}\right)^2}{\left(m_W^\mrm{SM}\right)^2}~~\textrm{for $\sqrt{s}\gg3.09~\TeV$}
\end{array}\right.\label{eq:Xiregimes}
\end{equation}
and $\Theta_{\sigma\Gamma}({s})$ is a %
coefficient of $\ILO{1}$ which is dependent on the energy scale and the physical process being corrected.
Parameter $\Xi_W(s)$ evaluates to
2~parts in $10^{3}$ in CDF~II and 4~parts in $10^{4}$ in ATLAS.} %

\subsubsection{\drafttwo{Specific examples}\label{sec:sxBRspecifics}}

\drafttwo{Currently, the highest-precision experimental values for cross-sections and decay constants are those obtained in CMS and ATLAS at energies of 8-13~$\TeV$,
corresponding to the domain $\sqrt{s}\gg 3.09~\TeV$. For the purpose of evaluating whether %
differences between CASMIR and the Standard Model yield testable modifications to %
these parameters, it is pragmatic to assume that comparable experimental precision could also be attained in the domain $m_W c^2\ll \sqrt{s}\ll 3.09~\TeV$. 
The value of the relative corrections $\Theta_{\sigma\Gamma}(s)\Xi_W(s)$ are therefore compared with the precision of the $13~\TeV$ measurements in both of the energy regimes considered in \Eref{eq:Xiregimes}.
The %
size of these corrections is also compared with published next-to-next-to-leading order (NNLO) Standard Model theoretical predictions for the same scenarios at $13~\TeV$.
Rather than going to the trouble of explicitly evaluating $\Theta_{\sigma\Gamma}(s)$ for a range of processes and energies, it is simpler to recognise that taking $\draft{|}\Theta_{\sigma\Gamma}(s)\draft{|}\rightarrow 10$ \draft{is expected to overestimate} the CASMIR corrections to the weak-boson-mediated scattering cross sections, and yet still produces corrections which are small compared with experimental and theoretical precision, as
shown in \tref{tab:sxBRshifts}.}

\chap{\begin{landscape}}\chapeight{\begin{landscape}}
\begin{table}
\tbl{\drafttwo{Magnitude of the CASMIR corrections %
to total cross-sections for specific Drell-Yan processes in the CASMIR low-energy (\prm{m_W c^2\ll\sqrt{s}\ll 3.09~\TeV}) and high-energy (\prm{\sqrt{s}\gg 3.09~\TeV}) regimes. Multiplication of the process-independent correction term \prm{\Xi_W(s)}~\peref{eq:Xiregimes} by a process-dependent factor \prm{\Theta_{\sigma\Gamma}(s)} yields the total correction; it may be reasonably assumed across all processes that this factor satisfies \prm{\draft{|}\Theta_{\sigma\Gamma}\draft{(s)|}<10}.
The experimental values shown here are %
high-precision measurements obtained at energies of 8-13~\prm{\TeV}. With larger discrepancies between CASMIR and the Standard Model being anticipated at lower energies, it is pragmatic to assume that %
comparable precision might similarly be obtained at energies \prm{m_W c^2\ll\sqrt{s}\ll 3.09~\TeV}.
Sources ATLAS, CMS15, and CMS16 represent experimental results reported in Refs.~%
\citen{aad2016,the-CMS-collaboration2015,the-CMS-collaboration2016} 
respectively, while NNLO represents the next-to-next-to-leading order theoretical results reported in \rcite{aad2016}. %
The mass of the Standard Model \prm{W}~boson is taken to be the updated ATLAS value of \prm{80.360(16)~\GeV/c^2}\notchap{.}\chap{~}\pcite{the-ATLAS-collaboration2023}\chap{.}\\
\draft{Regarding these corrections, for the theoretical values~(NNLO) at all energy scales and for the experimental datasets~(CMS15, CMS16, ATLAS) at \prm{m_Wc^2\ll\sqrt{s}\ll 3.09~\TeV} the values are the predicted discrepancies between CASMIR and the Standard Model. For the experimental datasets at \prm{\sqrt{s}\gg 3.09~\TeV} the values represent corrections due to differences between CASMIR and the Standard Model which, in CASMIR, are anticipated to already be present in the quoted experimental results.}
\\
Note: Processes \prm{\sigma(pp\rightarrow WX)\times\ldots} represent a sum over both charges of \prm{W} bosons, with emission of appropriate leptons.\\
\draft{Note 2: While the cross-sections for processes beginning \prm{\sigma(p\bar{p}\rightarrow WX)\times\ldots} differ from those beginning \prm{\sigma(pp\rightarrow WX)\times\ldots}, the relative magnitudes of the corrections to the cross-sections are predicted to be comparable.}
}}
{%
\drafttwo{
\chap{\begin{center}}\chapeight{\begin{center}}
\begin{tabular}{lcr@{.}lr@{~~}rr@{~~}r}%
\toprule
&&&&\multicolumn{2}{c}{\draft{Magnitude of c}orrection}&\multicolumn{2}{c}{\draft{Magnitude of c}orrection}\\
&&\multicolumn{2}{c}{Value}&\multicolumn{2}{c}{with $\draft{|}\Theta_{\sigma\Gamma}\draft{(s)|}=10$ at} & \multicolumn{2}{c}{with $\draft{|}\Theta_{\sigma\Gamma}\draft{(s)|}=10$ at}\\
\multicolumn{1}{c}{Total cross-section} & Source & \multicolumn{2}{c}{(\draft{in} nb)} & \multicolumn{2}{c}{$m_Wc^2\ll\sqrt{s}\ll 3.09~\TeV$} & \multicolumn{2}{c}{$\sqrt{s}\gg 3.09~\TeV$}\\
\midrule
$\sigma(pp\rightarrow W^+X)\times \mc{B}(W^+\rightarrow e^+\nu)$ & CMS15 
& ~\;11&39(65) & ~~~~~~~~~~$0.3\,\sigma$&$(\draft{210}~\mrm{pb})$~~~~&~~~~~~~$6\times 10^{-3}\,\sigma$&$(\draft{4}~\mrm{pb})$\\
$\sigma(pp\rightarrow W^+X)\times \mc{B}(W^+\rightarrow \mu^+\nu)$ & CMS15 
& 11&35(64) & $0.3\,\sigma$&$(\draft{210}~\mrm{pb})$~~~~&$6\times 10^{-3}\,\sigma$&$(\draft{4}~\mrm{pb})$\\
$\sigma(pp\rightarrow W^+X)\times \mc{B}(W^+\rightarrow \ell^+\nu)$ & CMS15 
& 11&37(60) & $0.4\,\sigma$&$(\draft{210}~\mrm{pb})$~~~~&$6\times 10^{-3}\,\sigma$&$(\draft{4}~\mrm{pb})$\\
$\sigma(pp\rightarrow W^+X)\times \mc{B}(W^+\rightarrow \ell^+\nu)$ & ATLAS 
& 11&83(41) & $0.5\,\sigma$&$(\draft{220}~\mrm{pb})$~~~~&$9\times 10^{-3}\,\sigma$&$(\draft{4}~\mrm{pb})$\\
$\sigma(pp\rightarrow W^+X)\times \mc{B}(W^+\rightarrow \ell^+\nu)$ & NNLO 
& 11&54(38) & $0.6\,\sigma$&$(\draft{210}~\mrm{pb})$~~~~&$1\times 10^{-2}\,\sigma$&$(\draft{4}~\mrm{pb})$\\
$\sigma(pp\rightarrow W^-X)\times \mc{B}(W^-\rightarrow e^-\bar\nu)$ & CMS15 
& 8&68(50) & $0.3\,\sigma$&$(\draft{160}~\mrm{pb})$~~~~&$6\times 10^{-3}\,\sigma$&$(\draft{3}~\mrm{pb})$\\
$\sigma(pp\rightarrow W^-X)\times \mc{B}(W^-\rightarrow \mu^-\bar\nu)$ & CMS15 
& 8&51(46) & $0.3\,\sigma$&$(\draft{160}~\mrm{pb})$~~~~&$6\times 10^{-3}\,\sigma$&$(\draft{3}~\mrm{pb})$\\
$\sigma(pp\rightarrow W^-X)\times \mc{B}(W^-\rightarrow \ell^-\bar\nu)$ & CMS15 
& 8&58(44) & $0.4\,\sigma$&$(\draft{160}~\mrm{pb})$~~~~&$6\times 10^{-3}\,\sigma$&$(\draft{3}~\mrm{pb})$\\
$\sigma(pp\rightarrow W^-X)\times \mc{B}(W^-\rightarrow \ell^-\bar\nu)$ & ATLAS 
& 8&79(30) & $0.5\,\sigma$&$(\draft{160}~\mrm{pb})$~~~~&$9\times 10^{-3}\,\sigma$&$(\draft{3}~\mrm{pb})$\\
$\sigma(pp\rightarrow W^-X)\times \mc{B}(W^-\rightarrow \ell^-\bar\nu)$ & NNLO 
& 8&54(27) & $0.6\,\sigma$&$(\draft{160}~\mrm{pb})$~~~~&$1\times 10^{-2}\,\sigma$&$(\draft{3}~\mrm{pb})$\\
$\sigma(pp\rightarrow WX)\times \mc{B}(W\rightarrow e\nu)$ & CMS15 
& 20&1(1.1) & $0.3\,\sigma$&$(\draft{370}~\mrm{pb})$~~~~&$6\times 10^{-3}\,\sigma$&$(\draft{7}~\mrm{pb})$\\
$\sigma(pp\rightarrow WX)\times \mc{B}(W\rightarrow \mu\nu)$ & CMS15 
& 19&9(1.1) & $0.3\,\sigma$&$(\draft{370}~\mrm{pb})$~~~~&$6\times 10^{-3}\,\sigma$&$(\draft{6}~\mrm{pb})$\\
$\sigma(pp\rightarrow WX)\times \mc{B}(W\rightarrow \ell\nu)$ & CMS15 
& 20&0(1.0) & $0.4\,\sigma$&$(\draft{370}~\mrm{pb})$~~~~&$6\times 10^{-3}\,\sigma$&$(\draft{6}~\mrm{pb})$\\
$\sigma(pp\rightarrow WX)\times \mc{B}(W\rightarrow \ell\nu)$ & ATLAS 
& 20&64(70) & $0.5\,\sigma$&$(\draft{380}~\mrm{pb})$~~~~&$1\times 10^{-2}\,\sigma$&$(\draft{7}~\mrm{pb})$\\
$\sigma(pp\rightarrow WX)\times \mc{B}(W\rightarrow \ell\nu)$ & NNLO 
& 20&08(65) & $0.6\,\sigma$&$(\draft{370}~\mrm{pb})$~~~~&$1\times 10^{-2}\,\sigma$&$(\draft{7}~\mrm{pb})$\\
$\sigma(pp\rightarrow ZX)\times \mc{B}(Z\rightarrow e^+e^-)$ & CMS15 
& 1&92(11) & $0.3\,\sigma$&$(\draft{35}~\mrm{pb})$~~~~&$6\times 10^{-3}\,\sigma$&$(\draft{0.6~\mrm{pb}})$\\
$\sigma(pp\rightarrow ZX)\times \mc{B}(Z\rightarrow \mu^+\mu^-)$ & CMS16 
& 1&870(62) & $0.6\,\sigma$&$(\draft{34}~\mrm{pb})$~~~~&$1\times 10^{-2}\,\sigma$&$(\draft{0.6~\mrm{pb}})$\\
$\sigma(pp\rightarrow ZX)\times \mc{B}(Z\rightarrow \ell^+\ell^-)$ & CMS15 
& 1&910(99) & $0.4\,\sigma$&$(\draft{35}~\mrm{pb})$~~~~&$6\times 10^{-3}\,\sigma$&$(\draft{0.6~\mrm{pb}})$\\
$\sigma(pp\rightarrow ZX)\times \mc{B}(Z\rightarrow \ell^+\ell^-)$ & ATLAS 
& 1&981(57) & $0.6\,\sigma$&$(\draft{37}~\mrm{pb})$~~~~&$1\times 10^{-2}\,\sigma$&$(\draft{0.6~\mrm{pb}})$\\
$\sigma(pp\rightarrow ZX)\times \mc{B}(Z\rightarrow \ell^+\ell^-)$ & NNLO 
& 1&890(66) & $0.5\,\sigma$&$(\draft{35}~\mrm{pb})$~~~~&$9\times 10^{-3}\,\sigma$&$~(\draft{0.6~\mrm{pb}})$\\
\botrule
\end{tabular}
\label{tab:sxBRshifts}
\chap{\end{center}}\chapeight{\end{center}}
}}
\end{table}%
\chap{\end{landscape}}\chapeight{\end{landscape}}

\drafttwo{In time, N$^3$LO calculations will improve the precision of the Standard Model prediction, permitting meaningful discrimination between the CASMIR and Standard Model values of Drell-Yan cross-sections. At present, however:
\begin{itemize}
\item NNLO calculations for Drell-Yan processes yield a precision of around 3\%~\cite{aad2016}. 
\item While N$^3$LO calculations have a lower
intrinsic uncertainty of around 2\% [denoted $\delta(\mrm{PDF}+\alpha_S)$%
], %
this is supplemented by an additional uncertainty arising because the PDF datasets used in the N$^3$LO Drell-Yan calculations have themselves only been extracted to NNLO precision [denoted $\delta(\textrm{PDF-TH})$]. %
The uncertainty of current N$^3$LO results is therefore also around 3-4\%~\cite{duhr2020,duhr2020a,duhr2022}. 
\end{itemize}
Thus, while present N$^3$LO calculations do improve on the central values for the Drell-Yan cross-sections, they do not as yet yield improvements in precision over the values given in \tref{tab:sxBRshifts}. Development of PDFs to N$^3$LO is ongoing~\cite{mcgowan2023,hekhorn2023}.}

\drafttwo{To summarise, the corrections to the massive weak vector boson production cross-sections %
arising from the CASMIR modifications to the electroweak sector are not presently %
accessible to experimental discrimination, being small compared both with experimental uncertainty and with the uncertainty in Standard Model calculations.}

\drafttwo{Finally, note that the corrections discussed here apply identically to all leptons, so %
these corrections do not imply any violation of lepton universality.}

\section{Muon $g-2$\label{sec:muong2}}

\subsection{Background and experimental status}

Measurement of the gyromagnetic anomaly of the muon also provides a highly sensitive test of the Standard Model. 
Classically, the gyromagnetic ratio of the muon is expected to take a value of~2. Quantum corrections adjust this value, with the corrected figure being written
\begin{equation}
g = 2(1+a_\mu)\qquad\mrm{or}\qquad g-2 = 2a_\mu
\end{equation}
where $a_\mu$ is the muon gyromagnetic anomaly. The Standard Model values of the electron and muon gyromagnetic anomalies have both been calculated to remarkable precision~\cite{aoyama2012,aoyama2018,aoyama2020}, and as probes of beyond-Standard-Model physics, the greater mass of the muon provides for greater sensitivity to corrections arising at the weak boson scale, and the theoretical values of these corrections in the Standard Model have been calculated to two loops and estimated to three~\cite{czarnecki2003,gnendiger2013}%
. Hadron terms are of similar magnitude, and \emph{de novo} calculation poses significant computational challenges, so the hadron contributions to the gyromagnetic anomaly are generally inferred from measurements on particle decay processes~\cite{aoyama2020}. The outcome of this formidable calculation is a prediction
\begin{align}
a_\mu^{\mrm{SM}}&=116591810(43)\times 10^{-11}\label{eq:aSM}
\end{align}
to which electroweak terms incorporating massive bosons (i.e.~$W$, $Z$, and Higgs) contribute
\begin{equation}
a_\mu^\mrm{EW}=153.6(1.0)\times 10^{-11}.\label{eq:aEWval}
\end{equation}
Curiously, this result is also in conflict with results from Fermilab. This time, the conflict is with the Muon~\mbox{$g-2$} experiment at Fermilab~\cite{aguillard2023a}
which in combination with the prior dataset from Brookhaven yields a global average of
\begin{equation}
a_\mu^\mrm{EXP}=1165920\draftone{59(22)}\times 10^{-11}.\label{eq:aEXP}
\end{equation}
At \draftone{$5.2\,\sigma$} tension this conflict \draftone{surpasses} the conventional $5\,\sigma$ tension required to declare a discovery. \draftone{However, there have been two important developments since the theoretical value $a^\mrm{SM}_\mu$ was obtained in 2020.  These are: %
\begin{itemize}
\item a lattice calculation of the Hadron Vacuum Polarisation (HVP) which disagrees with previous data-driven values obtained from $e^+e^-$ experiments~\cite{borsanyi2021}, and %
\item a measurement of the $e^+e^-\rightarrow \pi^+\pi^-$ cross-section at CMD-3 which also disagrees with previous $e^+e^-$ experiments.~\cite{ignatov2023} 
\end{itemize}
Attempts to reconcile these results are ongoing~\cite{colangelo2022}. Pending this reconciliation, it is likely that a contemporary re-evaluation of the Standard Model prediction for $a_\mu$ would carry greater uncertainty and thus smaller tension with experiment.}
Nevertheless, this result is also considered a strong candidate to indicate possible beyond-Standard-Model physics.

\subsection{Relevance to CASMIR and CDF~II}

It is possible to independently cross-check the predictions of CASMIR using the results of the Muon~\mbox{$g-2$} experiment, and doing so provides additional support for the CASMIR model of the CDF~II measurement of $W$~boson mass. Supplementing colorless photons, $W$,~and $Z$~bosons with their more numerous chromatic counterparts yields a correction to the value of $a_\mu^\mrm{EW}$ in CASMIR which is readily calculated at the 1-loop level and approximated at the 2-loop level. In principle the introduction of these species also affects the gluon/hadron contributions to $a_\mu$. However, since the values of these contributions are currently derived from experiment, the measured values are %
understood to already incorporate any contributions from all relevant physical processes (see also \sref{sec:hadronnote}). 

The resulting value of $a_\mu$
is shown in \sref{sec:muongyrocalc} to be
\begin{equation}
a_\mu^\mrm{CASMIR}=116592071(46)\times 10^{-11}\label{eq:aCASMIR}.
\end{equation}
Tension 
with experimental results is reduced from $\draftone{5.2}\,\sigma$ for the Standard Model to $0.2\,\sigma$ for CASMIR. %
This reduction %
in tension primarily reflects that the CASMIR central value differs from the experimental value by only \draftone{$0.5$} times %
the experimental uncertainty $\sigma^\mrm{EXP}$, compared with $\draftone{11}\,\sigma^\mrm{EXP}$ for the Standard Model.

\subsection{Calculation in CASMIR\label{sec:muongyrocalc}}

\subsubsection{Overview of massive vector boson exchange in CASMIR\label{sec:vecbosexchg}}

\paragraph{Bosons are chromatic\notchap{:}} As noted above, in the full $\SU{9}$ treatment of CASMIR %
the $W$ and $Z$ bosons are augmented by eight colored counterparts with masses $m_{W^{\tc}}=80.4434(22)~\GeV/c^2$ and $m_{Z^{\tc}}=91.\draftone{1928}(35)~\GeV/c^2$ respectively. These colored species appear in addition to the usual colorless bosons in color-rich environments (\sref{sec:chromenv}), and augment the masses of $W$~and $Z$ boson loops \schapnotchap{sec:tW}{pfeifer2022CASM4}. Similarly, the photon is supplemented by eight chromatic photons which are massless and have nontrivial representation on the $C$~sector.
Further, all fermion species in CASMIR are composed from three colored preons which are bound by gluon exchange, and all vector bosons are constructed from co-propagating preon/antipreon pairs. 

In \aref{apdx:massloops}%
\notchap{ of~\rcite{pfeifer2022CASM4}} it is argued that loop corrections to vertex functions in CASMIR are dominated by short-range contributions which may be identified as those on the order of half the wavelength of the loop boson(s) or less.
The muons in the Fermilab experiment are held in a storage ring at the ``magic momentum'' of $3.1~\GeV/c$, corresponding to an energy (including rest energy) of ${E}=3.2~\GeV$ prior to any fluctuations arising from interaction with the pseudovacuum. As discussed in \sref{sec:chromenv}%
, the timescale for comparison with pseudovacuum correlations is that associated with half a wavelength, %
\begin{equation}
\frac{1}{2}h\cdot[3.2~\GeV]^{-1} = h\cdot[6.4~\GeV]^{-1}.
\end{equation}
This timescale corresponds to an energy scale $\mc{E}=6.4~\GeV$ satisfying $\mc{E}_0<\mc{E}<\mc{E}_\Omega$~\eref{eq:Ebounds}, or equivalently the storage ring energy of $E=3.2~\GeV$ satisfies the relationship $1.79~\GeV<E<3.09~\TeV$~\eref{eq:Ebounds3}, permitting the appearance of 
\draftone{virtual} colored boson species.
\draftone{However, in contrast to the hadron collisions in CDF~II and ATLAS, the circulating muons do not experience a color-rich foreground environment and therefore there is no foreground chromatic $W$ or $Z$ boson formation. The role of these species is therefore restricted to appearances as short-lived virtual particles in loop corrections.}

\paragraph{Chromatic enhancement of photon emission\notchap{:}}

As discussed in \sref{sec:colouredAmass}\notchap{ of \rcite{pfeifer2022CASM4}}, the emission of chromatic photons may enhance the electromagnetic field of a charged particle where appropriate color freedoms exist. The gauge-like constraint that fermions always exhibit color neutrality \schapnotchap{sec:Csector}{pfeifer2022CASM4} prohibits the direct (tree-level) emission of chromatic photons from a lepton, but leaves open the prospect of their emission during electroweak loop corrections to the tree level process.

\begin{figure}
\includegraphics[width=\linewidth]{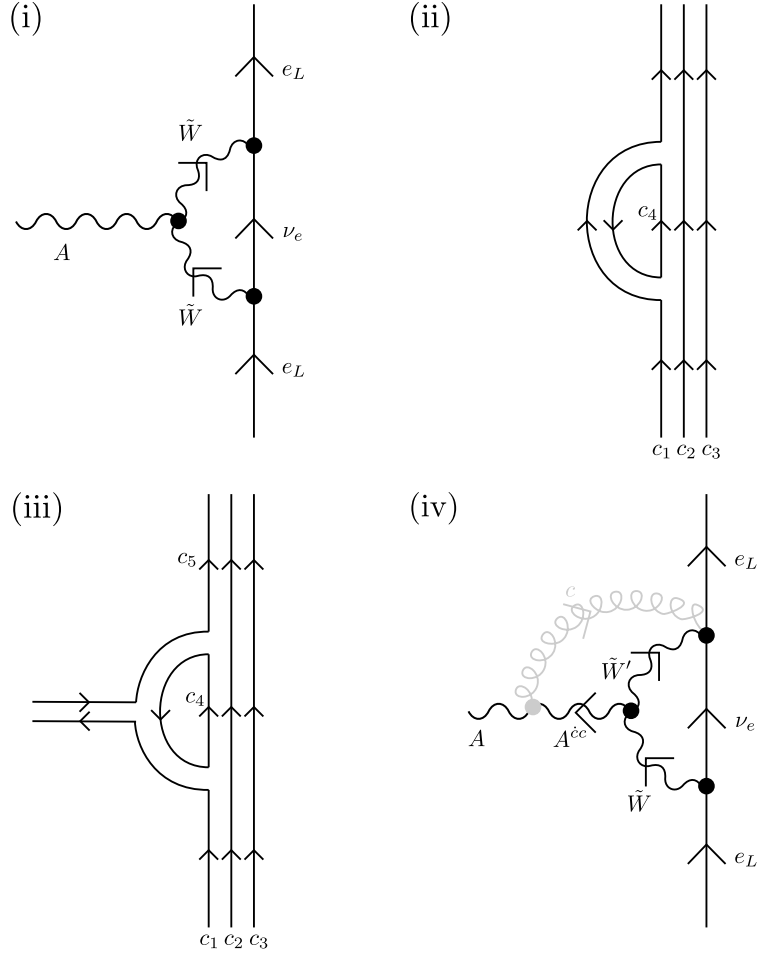}
\caption{(i)~One-loop \prm{W}~boson correction to the electromagnetic vertex. As per \draftone{\protect{\sref{sec:tW} of \rcite{pfeifer2022CASM4}}} the loop boson is denoted \prm{\tW}. (ii)~Color-labeled preon structure of the \prm{W}~boson loop. (iii)~Color-labeled preon structure of the \prm{W}~boson loop with photon emission. If the photon is permitted to be chromatic, the number of free preon color labels is increased by one. (iv)~One-loop \prm{W}~boson correction to the electromagnetic vertex, with chromatic photon emission in the \prm{e^C_{ij}} basis, and restoration of lepton color neutrality through a gluon-like change of \draftone{coordinate} frame. The notation \prm{\tW'} indicates that a color perturbation has acted on the superposition making up the loop \prm{W} boson.\label{fig:gluphotonemission}}
\end{figure}

Consider the photon emission processes shown in \fref{fig:gluphotonemission}.
In diagram~(i) a photon is emitted from a $W$~boson in a 1-loop correction to the electromagnetic vertex. The preon structure of the $W$~loop is shown in diagram~(ii), with photon emission being added in diagram~(iii). Comparing diagrams~(ii) and~(iii), if the photon is free to carry color charge then an additional independent color label is introduced for a multiplicative factor of three on the contribution of this loop. Separability of $A$ and $C$ sectors \achapnotchap{apdx:ACseparability}{pfeifer2022CASM4} ensures that chromatic photons contribute to the electromagnetic field in the far-field regime without needing to explicitly solve for confinement of the color charge \schapnotchap{sec:colouredAmass}{pfeifer2022CASM4}. In practice this process must simultaneously render the photon colorless and restore the color neutrality of the preon triplet making up the lepton; the latter effect is achieved by an induced change of \draftone{coordinate} frame as per \sref{sec:Csector} which only occurs when the $W$~boson is absorbed by the preon triplet. This \draftone{coordinate} change is functionally equivalent to exchange of a massless gluon with trivial vertices and a delta function over momentum. %
By color confinement and $A/C$~separability this may conveniently be assumed to recombine with the chromatic photon or its products and restore color neutrality as per \fref{fig:gluphotonemission}(iv), and by identification with a choice of \draftone{coordinate} frame %
the vertices and loop coefficient associated with this boson must evaluate to~1. %

Now recognize that the loop boson in diagram~(i) is as described in \sref{sec:tW}\notchap{ of \rcite{pfeifer2022CASM4}}, having a $\frac{2}{9}$ chance of being of definite coloration. When coloration is definite at source and sink, the preons at the $W^\tc W^{\tc\dagger}A$ vertex are constrained to be of identical color, and no factor of three accrues. For the other $\frac{7}{9}$, recognize that both preons are in equal superposition of color charges $r$, $g$, and $b$, and that summing across all elements of the $e^C_{ij}$ basis all color pairs are equally likely, up to those excluded as part of the $\frac{2}{9}$ for which color is fixed and definite. Further, by symmetry of muon construction under $\SU{3}_C$ the probability of any given coloration being associated with that $\frac{2}{9}$ fraction is equally distributed across all possible pairs. Interaction with the chromatic photons leaves the \draftone{components of} any $\SU{3}_C$-invariant initial state unchanged, and thus the $\frac{7}{9}$ fraction of loops which are not singly-labeled in the $e^C_{ij}$ basis may emit $\SU{3}_C$-invariant superpositions of chromatic photons while themselves remaining unchanged.

Finally, returning to the $\frac{2}{9}$ chance that the loop $W$ boson is of definite coloration, recognize that whereas the mass interaction of \sref{sec:tW}\notchap{ of \rcite{pfeifer2022CASM4}} takes place at multiple separated points along a finite interval which incorporates segments of fermion propagator, the electromagnetic interaction comprises a pointlike tree-level interaction and higher-order terms which are written as corrections to this point-like interaction. Corrections to the EM vertex must therefore explicitly incorporate factors arising from any fermion crossings which take place. In the mass interaction, the ability of the colored $W$ boson to be reabsorbed by either of two color-appropriate preons yielded an absorption probability of $\frac{2}{3}$ comprising $\frac{1}{3}+\frac{1}{3}$. These two terms differ by a preon pair exchange. In the mass interaction this exchange, if it occurs, is absorbed into the definition of the propagator segment. In the electromagnetic loop correction, on the other hand, the crossing yields a factor of~$-1$ which is not subsumed, and thus
\begin{equation}
\frac{1}{3}\times\left(\frac{1}{3}+\frac{1}{3}\right)=\frac{2}{9}\mrm{~~~becomes~~~}\frac{1}{3}\times\left(\frac{1}{3}-\frac{1}{3}\right)=0.
\end{equation}
Consequently any diagram emitting a photon from a loop $W$ boson attracts a net multiplicative factor across all terms of the $W$~boson superposition given by
\begin{equation}
1\times\left[\frac{1}{3}\times\left(\frac{1}{3}-\frac{1}{3}\right)\right]+3\times\left(7\times\frac{1}{3}\times\frac{1}{3}\right)=\frac{7}{3}.
\end{equation}
Denote this color-related factor by
\begin{equation}
\Theta_c:=\frac{7}{3}.\label{eq:Thetac}
\end{equation}
Note that %
discrimination between the $\frac{7}{9}$ and $\frac{2}{9}$ components of the $W$~boson superposition (and evaluation of their associated coefficients) is dependent on the structure of the diagram as a whole, and arises on a per-loop and not a per-vertex basis.

\begin{figure}
\includegraphics[width=\linewidth]{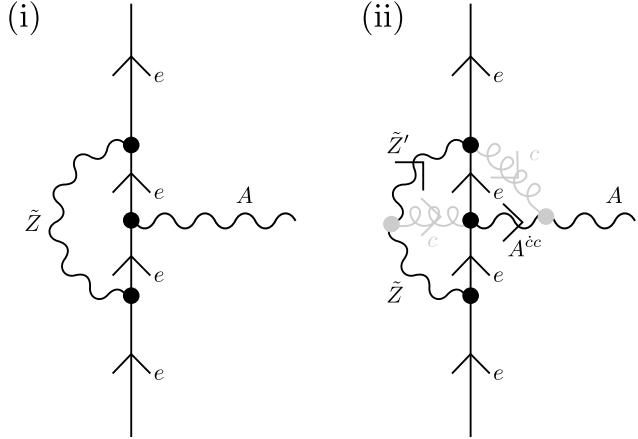}
\caption{ (i)~One-loop \prm{Z}~boson correction to the electromagnetic vertex. As per \protect{\schapnotchap{sec:tW}{pfeifer2022CASM4}} the loop boson is denoted \prm{\tilde{Z}}. (ii)~One-loop \prm{Z}~boson correction to the electromagnetic vertex, with chromatic photon emission in the \prm{e^C_{ij}} basis, and restoration of lepton color neutrality twice through gluon-like changes of \draftone{coordinate} frame. The notation \prm{\tilde{Z}'} indicates that a color perturbation has acted on the superposition making up the loop \prm{Z} boson~(\prm{\tilde{Z}}).\label{fig:gluphotonemissionZ}}
\end{figure}

Next, consider the $Z$~loop shown in \fref{fig:gluphotonemissionZ}(i). 
As with the loop $W$ boson, the preons within the $Z$~loop exist in a superposition of color and there is a freedom to act on these preons with color transformations which transform individual preons provided the superposition as a whole remains unchanged. If the photon coupling is replaced with a chromatic photon coupling, the interacting preon changes color. As this preon belongs to a fermion triplet, and such triplets define color neutrality as per \sref{sec:Csector}, neutrality must be immediately restored by a change of \draftone{coordinate} frame on $\Cw{18}$ which may be represented as a massless boson associated with a loop evaluating to a factor of~1.\footnote{In principle the process of color transfer may be arbitrarily complicated; however, since this gluon exchange is equivalent to a change in \draftone{coordinate} frame, there must always exist a choice of parameters in which the associated loop evaluates to a factor of 1.} In the presence of the $Z$~loop, this \draftone{coordinate} transform may be chosen to transfer the change of color to the $Z$~boson as shown in \fref{fig:gluphotonemissionZ}(ii). On recombination with the triplet, a further frame change carries this color transformation out to recombine with the color of the chromatic photon as discussed for the $W$~boson scenario above. Although the CASMIR loop $Z$ boson has the same mass as the free $Z$ boson, it nevertheless retains the same color considerations as the loop $W$ boson and therefore this color transfer is admissible seven times out of nine with the other two terms vanishing. Both $W$ and $Z$ loops therefore attract factors of $\Theta_c$.

\paragraph{Change in boson masses\notchap{:}}

The next factor to be considered is that associated with the $W$ boson mass.
Compared with the Standard Model scenario, $W$~boson masses in loops are rescaled by a factor
\begin{equation}
\Theta_m:=\frac{m_\tW^2}{m_W^2}\label{eq:Thetam}
\end{equation}
which evaluates in CASMIR as
\begin{equation}
\Theta_m=\frac{1+\frac{51}{18\left[k^{(e)}_{1}(\mc{E}_e)\,{N_0}\right]^4}}{1+\frac{19}{18\left[k^{(e)}_{1}(\mc{E}_e)\,{N_0}\right]^4}}+\OOOO{\left[k^{(e)}_{1}(\mc{E}_e)\,{N_0}\right]^{-8}}
\end{equation}
for
\begin{equation}
\left[k^{(e)}_{1}(\mc{E}_e)\,{N_0}\right]^{-4}=2.77905(21)\times 10^{-4}. %
\end{equation}
Since the present \paper{} evaluates $a_\mu^\mrm{CASMIR}$ as a correction to the result of \rcite{gnendiger2013}, the value of $\Theta_m$ is evaluated in \Eref{eq:Thetam} using the value of 
$m_W$ adopted in the original calculation, namely $80.363(13)~\GeV/c^2$.
The $Z$ and Higgs masses are not rescaled.

\subsubsection{1-loop terms}

The Standard Model one-loop corrections to the gyromagnetic ratio of the muon have been well understood for many decades~\cite{fujikawa1972,jackiw1972,altarelli1972,bars1972,bardeen1972}, incorporating contributions from the $W$, $Z$, and Higgs boson. However, for a Higgs boson of approximately $125~\GeV/c^2$ the Standard Model Higgs contribution is approximately $2.2\times 10^{-14}$~\cite{miller2012}, falling below the current threshold for experimental relevance. In CASMIR all Higgs vertices \draftone{%
below $\frac{1}{2}\mc{E}_\Omega\approx 3.1~\TeV$} are suppressed by a further factor of $\big[k^{(e)}_{1}(\mc{E}_e)\,{N_0}\big]^{-2}$ apiece, %
reducing this value to $6.1\times 10^{-18}$. The portion of the 1-loop corrections relevant at current experimental precision therefore takes the form
\begin{equation}
a_\mu^{\mrm{EW}(1)}=\frac{G_F m_\mu^2}{\sqrt 2\cdot8\pi^2}\left[\frac{10}{3}+\frac{1}{3}\left(1-4\sin^2\theta_W\right)^2-\frac{5}{3}\right]
\end{equation}
where the Fermi constant $G_F$ is proportional to $m_W^{\;-2}$. %
Within the square brackets, $10/3$ arises from the $W$~boson loop correction and the rest arises from the $Z$~boson loop correction.

As discussed in \sref{sec:vecbosexchg}, both the $W$~boson and $Z$~boson loop corrections are augmented by a factor of~$\Theta_c$ due to additional coupling to chromatic photons. 
Further, the value of $G_F$ is determined from muon decays and in CASMIR is therefore associated with the mass of the colorless $W$ boson. However, in loops the effective $W$~boson mass is given by $m_\tW^{\;-2}$ and the $W$~boson contributions to $a_\mu^\mrm{EW}$ must therefore be multiplied by a factor of $\Theta_m^{\;-1}$. There is no mass difference in CASMIR between free and loop $Z$~bosons and hence this factor accrues to the $W$~boson contributions only.

Finally, expanding 
\begin{equation}
\sin^2\theta_W=1-\frac{m_W^2}{m_Z^2},
\end{equation}
recognize that when $m_W^2$ appears in this expression it does so in the context of writing the $Z$~boson calculation in terms of the $W$~boson calculation---which was performed using the Standard Model $W$~\draftone{boson} mass, equivalent to the CASMIR achromatic $W$ \draftone{boson} mass. Therefore, no correction to the factors of $m_W^2$ appearing in $(1-4\sin^2\theta_W)^2$ are required.

The net correction is thus
\begin{align}
a_\mu^{\mrm{EW}(1)}\longrightarrow a_\mu^{\mrm{EW(1);CASMIR}}:=\,&\frac{\Theta_c G_F m_\mu^2}{\sqrt 2\cdot8\pi^2}\left[\frac{10}{3\,\Theta_m}+\frac{1}{3}\left(1-4\sin^2\theta_W\right)^2-\frac{5}{3}\right]\\
=\,&\Theta_c\Theta_M^{(1)}\, a_\mu^{\mrm{EW}(1)}
\end{align}
where
\begin{equation}
\Theta_M^{(1)}:=\frac{\frac{10}{3\,\Theta_m}+\frac{1}{3}\left(1-4\sin^2\theta_W\right)^2-\frac{5}{3}}{\frac{10}{3}+\frac{1}{3}\left(1-4\sin^2\theta_W\right)^2-\frac{5}{3}}.
\end{equation}
This increases the value of $a_\mu^{\mrm{EW}(1)}$ from $194.80(1)\times 10^{-11}$ to $454.18(30)\times 10^{-11}$,
with the factor $\Theta_M^{(1)}$ contributing $-0.35(30)\times 10^{-11}$.

\subsubsection{2-loop terms and beyond}

Whereas it is relatively straightforward to calculate the 1-loop corrections in CASMIR, full evaluation of the 2-loop corrections is substantially more involved and is beyond the scope of this %
\paper{}. However, following \rcite{gnendiger2013} to write
\begin{align}
\begin{split}
a_\mu^\mrm{EW}=\;&a_\mu^{\mrm{EW}(1)}+a^{\mrm{EW}(2)}_{\mu;\mrm{bos}}+a^{\mrm{EW}(2)}_{\mu}(e,\mu,u,c,d,s)+a^{\mrm{EW}(2)}_{\mu}(\tau,t,b)\\
&+a^{\mrm{EW}(2)}_{\mu;\textrm{f-rest,H}}+a^{\mrm{EW}(2)}_{\mu;\textrm{f-rest,no\,H}}+a^{\mrm{EW}(\geq 3)}_{\mu},
\end{split}\label{eq:aEWsplit}
\end{align}
a reasonable estimate of the correction may be obtained term-by-term at the cost of only a relatively small increase in uncertainty. %
In keeping with \rcite{gnendiger2013}, uncertainties in these terms are treated as correlated. 
There are two relevant sources of corrections to these terms. First, chromatic bosons may introduce factors of $\Theta_c$. Second, terms involving Higgs exchange are suppressed by factors of $\big[k^{(e)}_{1}(\mc{E}_e)\,{N_0}\big]^{-4}$. 
There are also corrections due to replacement of the loop $W$ boson mass $m_W^{\;-2}\rightarrow m_\tW^{\;-2}$, but with \draftone{this contribution to} the 1-loop correction already right on the edge of significance, it is assumed that \draftone{the corresponding contributions to} the 2-loop corrections may be safely neglected.

Begin with the corrections due to colored bosons. In diagrams containing chromatic $W$ and $Z$ bosons, photon emission may be augmented by the color sector. However, such diagrams must not contain a color ``tadpole''. If color enters a loop on one particle only, then by conservation of charge on the $C$~sector the color associated with that particle must be trivial. \chap{Equivalent statements are that
\begin{itemize}
\item the charge loop of Eq.~(2.7) of \rcite{bonderson2008} yields a vanishing delta function;
\item there exists no way to assign non-vanishing generalized Clebsch-Gordan coefficients to these diagrams unless the color on the single leg vanishes.
\end{itemize}}
By \rcitet{aoyama2020},~\citen{keshavarzi2022},~\citen{gnendiger2013}, and \citen{gribouk2005} the majority of two-loop diagrams which admit colored bosons under this criterion appear in term $a^{\mrm{EW}(2)}_{\mu;\mrm{bos}}$. Some diagrams in this term attract a factor of $\Theta_c$ and some do not. 
Therefore introduce
\begin{equation}
\Theta_A^r:=1+\frac{\Theta_c-1}{2}\pm\frac{\Theta_c-1}{2}
\end{equation}
where the superscript $^r$ denotes that the values of $\Theta_A^r$ mark the extrema of a range from~1 to $\Theta_c$.
Some diagrams in $a^{\mrm{EW}(2)}_{\mu;\mrm{bos}}$ also involve Higgs exchange, therefore similarly introduce
\begin{align}
\Theta_\bmh^r&:= f_\bmh + \frac{1-f_\bmh}{2}\pm\frac{1-f_\bmh}{2}\\
f_\bmh&:=\big[k^{(e)}_{1}(\mc{E}_e)\,{N_0}\big]^{-4}.
\end{align}
The CASMIR value of $a^{\mrm{EW}(2)}_{\mu;\mrm{bos}}$, denoted $a^{\mrm{EW}(2);\mrm{CASMIR}}_{\mu;\mrm{bos}}$, then falls within the range bounded by the extrema of $\Theta_A^r\Theta_\bmh^r\,a^{\mrm{EW}(2)}_{\mu;\mrm{bos}}$.

\begin{figure}
\includegraphics[width=\linewidth]{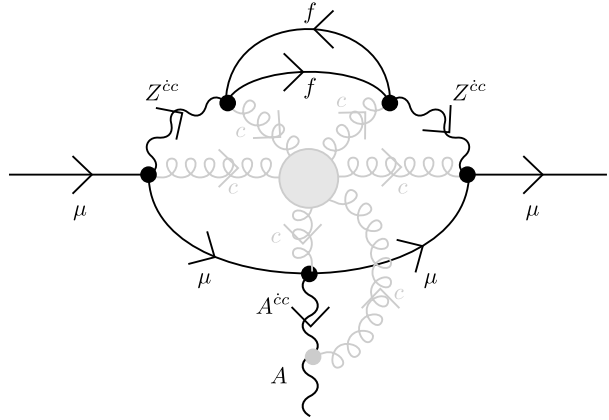}
\caption{\prm{ZZ\gamma} term of \prm{a^{\mrm{EW}(2)}_{\mu;\textrm{f-rest,no\,H}}} with chromatic \prm{Z} bosons in the \prm{e^C_{ij}} basis, with the induced \draftone{coordinate} frame change on \prm{\Cw{18}} represented by the gluons in grey. When photon emission is chromatic this also is associated with an induced gluon interaction to restore color neutrality on the muon, and the chromatic photon likewise returns an induced gluon on reducing to an achromatic photon over distances large compared with the strong scale. It is unnecessary to evaluate the contents of the shaded circle as the color redistribution process as a whole corresponds to nothing more than a change of \draftone{coordinate} frame on \prm{\Cw{18}}, and therefore yields a numerical factor of~1. Adapted from Fig.~1(e) of \prcite{czarnecki1995}.\label{fig:ThetacZZterm}}
\end{figure}

It is convenient to
replace $\Theta_A^r$ and $\Theta_\bmh^r$ with estimated values $\Theta_A$ and $\Theta_\bmh$ having an uncertainty derived from the indicated ranges. 
Given full knowledge of the two-loop corrections this value could be evaluated precisely, excepting only the uncertainties arising from the input parameters, and \draftone{up to these uncertainties%
} the distributions associated with $\Theta_A$ and $\Theta_\bmh$ would be the Dirac delta function. 
On the other hand, assuming no prior knowledge of the relative frequencies of the %
boson species in the loop corrections %
yields the most %
pessimistic %
unimodal distribution, being one which is flat between the upper and lower bounds and zero outside. For such a distribution, 68.3\% of values lie within 68.3\% of the range. Therefore a maximally conservative approximation is obtained by defining
\begin{align}
\Theta_A&:=1+\frac{\Theta_c-1}{2}\\
\Delta\Theta_A&:=\frac{\Theta_c-1}{2}\times0.683\\
\Theta_\bmh&:=f_\bmh + \frac{1-f_\bmh}{2}\\
\Delta\Theta_\bmh &:= \frac{1-f_\bmh}{2}\times0.683
\end{align}
and treating $\Delta\Theta_A$ and $\Delta\Theta_\bmh$ as the effective $1\,\sigma$ uncertainties of $\Theta_A$ and $\Theta_\bmh$ respectively. 
The value of $a^{\mrm{EW}(2);\mrm{CASMIR}}_{\mu;\mrm{bos}}$ is then given by
\begin{equation}
\begin{split}
a^{\mrm{EW}(2);\mrm{CASMIR}}_{\mu;\mrm{bos}}:=\,&\Theta_A\Theta_\bmh\,a^{\mrm{EW}(2)}_{\mu;\mrm{bos}}\\
=\,&-17(12)\times 10^{-11}.
\end{split}
\end{equation}
Three other terms differ between CASMIR and the Standard Model. These are $a^{\mrm{EW}(2)}_{\mu;\textrm{f-rest,H}}$, which always contains a Higgs boson and therefore in CASMIR attracts a factor of $f_\bmh$, 
\begin{equation}
\begin{split}
a^{\mrm{EW}(2);\mrm{CASMIR}}_{\mu;\textrm{f-rest,H}}:=\,&f_\bmh\, a^{\mrm{EW}(2)}_{\mu;\textrm{f-rest,H}}\\
=\,&-0.0004196(28)\times 10^{-11},
\end{split}
\end{equation}
$a^{\mrm{EW}(2)}_{\mu;\textrm{f-rest,no\,H}}$, which attracts a factor of $\Theta_c$ on its $ZZ\gamma$ term only, as shown in \fref{fig:ThetacZZterm},
\begin{equation}
\begin{split}
a^{\mrm{EW}(2);\mrm{CASMIR}}_{\mu;\textrm{f-rest,no\,H}}:=\,&\Theta_A\, a^{\mrm{EW}(2)}_{\mu;\textrm{f-rest,no\,H}}\\
=\,&-7.7(2.1)\times 10^{-11},
\end{split}
\end{equation}
and $a^{\mrm{EW}(\geq 3)}_{\mu}$ which is an uncertainty attributed to three-loop terms. For $a^{\mrm{EW}(\geq 3);\mrm{CASMIR}}_{\mu}$ assume the greatest possible increase in uncertainty,
\begin{equation}
\begin{split}
a^{\mrm{EW}(\geq 3);\mrm{CASMIR}}_{\mu}:=\,&\Theta_c\, a^{\mrm{EW}(\geq 3)}_{\mu}\\
=\,&0.00(47)\times 10^{-11}.
\end{split}
\end{equation}
All corrections to $a^{\mrm{EW}(2)}_{\mu}$ and $a^{\mrm{EW}(\geq 3)}_{\mu}$ are summarized in \tref{tab:EW2factors}.
\begin{table}
\tbl{Estimated corrections to two-loop and three-loop contributions to the muon gyromagnetic anomaly in CASMIR. These corrections arise due to emission of chromatic photons, a correction to the \prm{W}~boson mass in electroweak loops, and the relative suppression of Higgs vertices in CASMIR. Standard Model values are as per \prcite{gnendiger2013}, with \prm{a^{\mrm{EW}(2)}_{\mu;\mrm{bos}}} and \prm{a^{\mrm{EW}(2)}_{\mu;\textrm{f-rest,H}}} updated as per \prcite{aoyama2020}.}
{
\chap{\begin{center}}\chapeight{\begin{center}}
\begin{tabular}{@{}lllrr@{}}
\toprule
Term&Boson species&Relative factor&SM value $\times 10^{11}$&CASMIR value $\times 10^{11}$\\
\midrule
$a^{\mrm{EW}(2)}_{\mu;\mrm{bos}}$ & $W/Z,~W/Z/A/H$ & $\Theta_A\Theta_\bmh$ & $-19.96(1)$ & $-17(12)$\\
$a^{\mrm{EW}(2)}_{\mu}(e,\ldots)$ & $Z,~A$ & $1$ & $-6.91(50)$ & $-6.91(50)$\\
$a^{\mrm{EW}(2)}_{\mu}(\tau,\ldots)$ & $Z,~A$ & $1$ & $-8.21(10)$ & $-8.21(10)$\\
$a^{\mrm{EW}(2)}_{\mu;\textrm{f-rest,H}}$ & $H,~Z/A$ & $f_\bmh$ & $-1.51(1)$ & $-0.0004196(28)$\\
$a^{\mrm{EW}(2)}_{\mu;\textrm{f-rest,no\,H}}$ & $W/Z,~W/Z/A$ & $\Theta_A$ & $-4.64(10)$ & $-7.7(2.1)$\\
$a^{\mrm{EW}(\geq 3)}_{\mu}$ & Any & $\Theta_c$ & $0.00(20)$ & $0.00(47)$\\
\botrule
\end{tabular}\label{tab:EW2factors}
\chap{\end{center}}\chapeight{\end{center}}
}
\end{table}%

Collecting all terms and assuming fully correlated errors yields
\begin{equation}
\begin{split}
a^{\mrm{EW};\mrm{CASMIR}}_\mu=415(16)\times 10^{-11}.\label{eq:aEWCASMIR}
\end{split}
\end{equation}

\subsubsection{A note on hadronic contributions\label{sec:hadronnote}}

In addition to the electromagnetic and electroweak terms, the gyromagnetic anomaly also receives significant contributions from the hadronic corrections denoted $a^\mrm{HVP}_\mu$ and $a^\mrm{HLbL}_\mu$~\cite{keshavarzi2022,aoyama2020}. In a first-principles calculation, these terms would also admit corrections due to the introduction of chromatic $W$ and $Z$~bosons on energy scales between $\mc{E}_0$ and $\mc{E}_\Omega$. However, these first-principles calculations have not yet been achieved to the level of precision required, and consequently the values of $a^\mrm{HVP}_\mu$ and $a^\mrm{HLbL}_\mu$ are evaluated using data-driven techniques. %
If CASMIR provides either a good description of reality (the chromatic $W$ and $Z$~boson are real) or a good analogue to experiment (other behaviors occur whose effects on $a_\mu$ are approximated by the chromatic bosons), then the processes which generate input data for these data-driven calculations already incorporate any necessary effects attributed in CASMIR to the chromatic weak bosons, assuming that this input data also comes from color-rich environments, e.g.~colliders operating at appropriate energy scales.

Consequently, in the current data-driven approach no corrections are needed to the values of $a^\mrm{HVP}_\mu$ and $a^\mrm{HLbL}_\mu$ adopted in the final calculations of \rcites{aoyama2020}{keshavarzi2022}. Corrections involving chromatic weak bosons would, however, be anticipated to be required for first-principle\draftone{s} calculations---again see \rcitet{aoyama2020},~\citen{keshavarzi2022}, and references therein. %

\subsubsection{Total muon gyromagnetic anomaly in CASMIR\label{sec:totalamu}}

Incorporating the CASMIR value for $a^{\mrm{EW}}_\mu$ as per \Eref{eq:aEWCASMIR} yields
\begin{equation}
a_\mu^\mrm{CASMIR}= 116592071(46)\times 10^{-11}. \tagref{eq:aCASMIR} %
\end{equation}
Despite the approximations employed, this computed result is %
only fractionally less precise than the Standard Model figure of
$a_\mu^{\mrm{SM}}=116591810(43)\times 10^{-11},%
$ and in contrast with the Standard Model result, is in near-perfect agreement with experiment (tension %
$0.2\,\sigma$ with BNL/Fermilab combined value; see \tref{tab:muonresults}).  %
\begin{table}
\tbl{Comparison of muon gyromagnetic anomaly results from %
combined Fermilab and BNL data~(\prm{a^\mrm{EXP}_\mu}) with predictions from the Standard Model~(\prm{a^\mrm{SM}_\mu}) and CASMIR~(\prm{a^\mrm{CASMIR}_\mu}). Differences are expressed as tensions and as multiples of the experimental uncertainties. %
Values smaller than \prm{1\,\sigma} are displayed in green online.}
{
\chap{\begin{center}}\chapeight{\begin{center}}
\begin{tabular}{l@{$\qquad\qquad\qquad$}c@{$\qquad\qquad$}c}
\toprule
&Tension \draftone{with $a_\mu^\mrm{EXP}$}&\multicolumn{1}{l}{Difference as a multiple of $\sigma^\mrm{EXP}$}\\
\midrule
$a^\mrm{SM}_\mu$&$\draftone{5.2}\,\sigma$&$\draftone{\p{.3}11}\,\sigma$\\
$a^\mrm{CASMIR}_\mu$&\textcolor{darkgreen}{$0.2\,\sigma$}&\textcolor{darkgreen}{$\draftone{\p{1}0.5}\,\sigma$}\\
\botrule
\end{tabular}\label{tab:muonresults}
\chap{\end{center}}\chapeight{\end{center}}
}
\end{table}%

There is scope for future calculations to 
further reduce the uncertainty in the CASMIR value by fully evaluating the two-loop electroweak terms using the colored $W$ and $Z$ bosons \draftone{alongside} their achromatic counterparts.

\section{Conclusion}

The Classical Analogue to the Standard Model In pseudo-Riemannian spacetime (CASMIR) effectively reproduces many properties of the Standard Model. In addition, it predicts the higher $W$~boson mass detected by CDF~II with a tension of $0.1\,\sigma$. [Observed: $80.4335(94)~\GeV/c^2$. Predicted: $80.4340(22)~\GeV/c^2$.] In CASMIR this increase in mass arises due to the electroweak vector bosons sometimes being able to carry color charges, with the %
\draftone{heavier,} colored $W$ \draftone{boson mass vertices} being excluded from the ATLAS calculation but %
\draftone{contributing} in CDF~II. This mechanism also alters the value of the muon gyromagnetic anomaly, yielding a value more consistent with experiment (tension reduced from $\draftone{5.2}\,\sigma$ to $0.2\,\sigma$).

The utility of an analogue model is determined by its ability to make predictions about experimental results. In addition to the \draftone{shift in apparent $W$~boson mass, which occurs only in hadron collisions where $\sqrt{s}$ is small compared with $3.09~\TeV$}, %
CASMIR also predicts %
\draftone{a shift in apparent $Z$~boson mass under the same conditions, increasing from $91.1877(35)~\GeV/c^2$ to $91.1922(35)~\GeV/c^2$. 
Existing experimental data lacks the precision to definitively probe this small ($4.5~\MeV/c^2$) correction.
}

\draftone{Current and planned future LHC beam energies are large when compared with the CASMIR chromatic mass vertex suppression threshold of $\sqrt{s}<
3.09~\TeV$, so future measurements at the LHC are anticipated to yield the lower (achromatic) values for both $W$~boson and $Z$~boson mass.
Nevertheless, it is feasible to probe this mass augmentation effect using existing and near-term forthcoming apparatus. The optimal configuration would be a high-luminosity hadron collider such as the LHC or the proposed HL-LHC upgrade~\cite{apollinari2015}, operating sufficiently below the chromatic mass vertex suppression threshold of
$3.09~\TeV$. %
(The former Tevatron accelerator satisfied this condition with $\sqrt{s}=1.96~\TeV$.)
}
\draftone{Appropriate lower-energy runs could be performed during the commissioning process for the HL-LHC.
}

Finally, %
if the domain of utility for CASMIR is sufficiently broad then it is %
possible that %
observations in astroparticle physics 
might %
support %
or exclude higher-generation weak bosons at the energies predicted in \tref{tab:heavyweak}, or that
some dark matter detection program or study of scalar boson interactions at the LHC might %
suggest existence of a species consistent with \sref{sec:DM}. 
However, there is no imminent expectation for existing experiments to %
test %
these latter predictions.

\notchap{
\section*{Acknowledgments}

This research was supported in part by the Perimeter Institute for Theoretical Physics.
Research at the Perimeter Institute is supported by the Government of Canada through Industry Canada and by the Province of Ontario through the Ministry of Research and Innovation.
The author thanks the Ontario Ministry of Research and Innovation Early Researcher Awards (ER09-06-073) for financial support.
This project was supported in part through the Macquarie University Research Fellowship scheme.
This research was supported in part by the ARC Centre of Excellence in Engineered Quantum Systems (EQuS), Project No.~CE110001013.
}
    
\appendix

\section{Mass relations from CASMIR\label{apdx:massrel}}

The geometry of CASMIR fixes a number of relationships between particle masses %
in the analogue model. The values of $\alpha$, $m_e$, and $m_\mu$ may be taken as input, with the following expressions
then yielding derived values
for $m_W$, $m_Z$, $m_\bmh$, $m_\tau$, %
$N_0$, and $m_{\tW}$ (with the latter two being internal parameters of the CASMIR model):
\begin{align}
&%
\frac
{m_e^2\left[k^{(e)}_i(\mc{E}_{e_i})\right]^4\left[1+\Delta_e(m_{e_i},\mc{E}_{e_i})\right]}
{m_{e_i}^2\left[k^{(e)}_1(\mc{E}_{e_i})\right]^4\left[1+\Delta_e(m_e,\mc{E}_{e_i})\right]}=1+\mc{O}_e(m_{e_i},\mc{E}_{e_i})%
\label{eq:test1}
\\
\nn&\frac
{9m_{\bmh}^2\left[1+\left(64+\frac{3}{2\pi}-f_Z\right)\frac{\alpha}{2\pi}\right]}
{20m_\tW^2\left\{\left(1-\frac{2}{3N_0}+\frac{1}{3{N_0}^2}\right)\left[1+\frac{30\alpha}{9\pi}\left(1+\frac{1}{3N_0}\right)\right]+\frac{1}{2\pi}\left[1+\frac{30\alpha}{\pi}\left(1-\frac{1}{3N_0}\right)\right]\right\}}\\
&\qquad\qquad=\frac{1+\frac{39}{18\left[k^{(e)}_{1}(\mc{E}_e)\,{N_0}\right]^4}}{1+\frac{51}{18\left[k^{(e)}_{1}(\mc{E}_e)\,{N_0}\right]^4}}(1+\mc{O}_b)
\label{eq:test2}
\\
&\frac
{3m_Z^2\left[1+\left(64+\frac{3}{2\pi}-f_Z\right)\frac{\alpha}{2\pi}\right]\left\{1+\frac{51}{18\left[k^{(e)}_{1}(\mc{E}_e)\,{N_0}\right]^4}\right\}}
{4m_\tW^2\left[1+\left(\frac{401}{12}+\frac{3}{2\pi}\right)\frac{\alpha}{2\pi}\right]\left\{1+\frac{55}{18\left[k^{(e)}_{1}\,(\mc{E}_e)\,{N_0}\right]^4}\right\}}=1+\mc{O}_b\label{eq:test3}
\\
&\frac
{18{N_0}^{4}m_e^2\Biggl[1+\left(64+\frac{3}{2\pi}-f_Z\right)\frac{\alpha}{2\pi}\Biggr]\left\{1+\frac{51}{18\left[k^{(e)}_{1}(\mc{E}_e)\,{N_0}\right]^4}\right\}}
{m_\tW^2\left(1+\frac{2}{N_0}\right)^{4}\left(1+\frac{1}{N_0}\right)^{4}\left[1+\Delta_{e}(m_e,\mc{E}_e)\right]}=1+\mc{O}_b+\mc{O}_e(m_e,\mc{E}_e)%
\\
&\frac{m_W^2\left\{1+\frac{51}{18\left[k^{(e)}_{1}(\mc{E}_e)\,{N_0}\right]^4}\right\}}{m_\tW^2\left\{1+\frac{19}{18\left[k^{(e)}_{1}(\mc{E}_e)\,{N_0}\right]^4}\right\}}=
1+\mc{O}_b\label{eq:VIII:mWratio}
\end{align}
with $e_i\in\{\mu,\tau\}$ corresponding to $i\in\{2,3\}$ respectively in \Eref{eq:test1}.\\
Additional symbols appearing in the above expressions are defined as follows:
\begin{align}
\nn\Delta_{e}(m_{e_i},\mc{E}):=\,&
\frac{8\alpha}{3N_0(3\alpha+2\pi)}\left\{1+\frac{(10\pi+180\alpha)m_{e_i}^2}{3\pi\left[m_c^*(\mc{E})\right]^2}+\frac{(5-4f_Z)\alpha m_{e_i}^2}{4\pi m_\tW^2}\right\}%
+\frac{90\alpha m_{e_i}^2}{\pi\left[m_c^*(\mc{E})\right]^2}
\\
\nn&+\frac{(5-4f_Z)\alpha m_{e_i}^2}{2\pi m_\tW^2}+\frac{5m_{e_i}^2}{[m_c^*(\mc{E})]^2}\left\{1+\frac{90\alpha m_{e_i}^2}{\pi\left[m_c^*(\mc{E})\right]^2}+\frac{(25-12f_Z)\alpha m_{e_i}^2}{6\pi m_\tW^2}\right\}\\
&+\frac{40m_{e_i}^2}{3m_{\bmh}^2\left[k^{(e)}_{1}(\mc{E}_e)\,{N_0}\right]^4}
+\mc{O}_e(m_{e_i},\mc{E})
\end{align}
\begin{align}
\begin{split}
\theta_e(\mc{E}) := -&\frac{3\pi}{4}\bm{\Biggl(}1-\frac{4\sqrt{%
\delta_e\{r[\Delta_\taue(m_\tau,\mc{E})-\Delta_\taue(m_\mu,\mc{E})]\}}}{3\pi}\bm{\Biggr)}
\\&\times\bm{\left(}1+\frac{4}{3\pi}\sqrt{\delta_e\left\{r\left[\frac{1+\Delta_\taue(m_e,\mc{E})}{1+\Delta_\taue(0,\mc{E})}-1\right]\right\}}\bm{\right)}
\end{split}
\end{align}
\begin{align}
m_c^2&:=m_\tW^2\left\{\frac
{1+\frac{131}{18\left[k^{(e)}_{1}(\mc{E}_e)\,{N_0}\right]^4}}
{1+\frac{51}{18\left[k^{(e)}_{1}(\mc{E}_e)\,{N_0}\right]^4}}\right\}
\left(1+\mc{O}_b\right)\\
{\left[m_c^*(\mc{E})\right]^2} &:= m_c^2\left(1-\frac{27}{10}\frac{\mc{E}^2}{m_c^2c^4}\right)
\\
f_Z&:=\frac{4}{3}\left(1-\frac{6m_\tW^2}{m_Z^2}+\frac{4m_\tW^4}{m_Z^4}\right)
\\
\mc{E}_\ell &:= m_\ell c^2
\\
r(n) &:= n\cdot \sqrt{1-\frac{n}{9}\;}
\\
k^{(\ell)}_n(\mc{E}) &:= 1+\sqrt{2}\cos{\left[\theta_\ell(\mc{E})-\frac{2\pi(n-1)}{3}\right]}
\\
\delta_e(n) &:= \sqrt{1+\frac{\pi^2n}{8}\left(1+\frac{\pi^2n}{32}\right)+\ILO{n^3}}~-1\label{eq:delta}
\\
\mc{O}_b&:=\OOOO{\frac{\alpha}{\pi}\left[k^{(e)}_1(\mc{E}_e)\,{N_0}\right]^{-4}}+\OO{\frac{\alpha^2}{\pi^2}}\label{eq:VIII:Ob}
\end{align}
\begin{align}
\nn\mc{O}_e(m_{e_i},\mc{E})
:=\,&\OO{\frac{\alpha}{\pi{N_0}^2}}+\OO{\frac{\alpha^2}{\pi^2{N_0}}}+\OOO{\frac{\alpha m_{e_i}^4}{\pi N_0[m_c^*(\mc{E})]^4}}
+\OOOO{\frac{\alpha^2 m_{e_i}^2}{\pi^2\left[m_c^*(\mc{E})\right]^2}}
\\&
+\OOOO{\frac{m_{e_i}^6}{\left[m_c^*(\mc{E})\right]^6}}
+\OOOO{\frac{\alpha m_{e_i}^2}{\pi m_\bmh^2\left[k^{(e)}_1(\mc{E}_e)\,{N_0}\right]^4}}\label{eq:VIII:Oe}\\&
+\OOOO{\frac{m_{e_i}^2}{m_\bmh^2\left[k^{(e)}_1(\mc{E}_e)\right]^4{N_0}^5}}
+\OOOO{\frac{m_{e_i}^4}{m_\bmh^4\left[k^{(e)}_1(\mc{E}_e)\,{N_0}\right]^4}}\nn
\end{align}

Noting that \Erefr{eq:test1}{eq:VIII:mWratio} %
yield six independent relationships, as \Eref{eq:test1} may be evaluated for both $i=2$ and $i=3$, 
the derived values for particle %
masses %
in the CASMIR model %
are given in \tref{tab:masses}.
\begin{table}[tb]
\tbl{Particle masses and coupling constants in CASMIR: Observed values, values predicted by CASMIR, and values predicted by the Standard Model (where these exist). 
Parameters \prm{N_0} and \prm{m_\tW} are also fixed by the equations of \pAref{apdx:massrel}, and evaluate to %
\prm{N_0=191.9470(37)} and \prm{m_\tW=80.3786(22)~\GeV/c^2} respectively. Mass uncertainties in CASMIR are estimates, so tensions in this table are given as an upper bound obtained by taking the CASMIR uncertainty to zero. %
}
{
\chap{\small}
\chap{\begin{center}}\chapeight{\begin{center}}
\begin{tabular}{lrr@{\chap{~}}lrcrc} %
\toprule
\multicolumn{2}{c}
{Parameter}&
\multicolumn{2}{c}{Observed value} &\multicolumn{2}{c}{\p{$^{21-24}$}CASMIR \notchap{\pcite{pfeifer2022CASM1,pfeifer2022CASM2,pfeifer2022CASM3,pfeifer2022CASM4}}}&\multicolumn{2}{c}{Standard Model}\\
&&&&& Upper bound &&\\
&&&& \multicolumn{1}{c}{Calculated} & %
on tension $(\sigma)$ &\multicolumn{1}{c}{Calculated}&Tension $(\sigma)$%
\\\midrule
$m_\tau$&$(\MeV/c^2)$& 1776.86(12)&\pcite{workman2022}%
~~& 1776.867413(43)~ & $\p{<}\;0.1$ & ---\p{$^*$} & ---\\ %
$m_W$&$(\GeV/c^2)$& 80.360(16)&\pcite{the-ATLAS-collaboration2023}%
~~& 80.3587(22)~ & $\p{<}\;0.1$ & 80.356(6) %
{\pcite{workman2022,awramik2004,erler2019}}\p{$^*$} & 0.2\\ %
$m_Z$&$(\GeV/c^2)$& 91.1876(21)&\pcite{workman2022}%
~~& 91.1877(35)~ & $<0.1$ & ---\p{$^*$} & ---\\ %
$m_\bmh$&$(\GeV/c^2)$& 125.11(11)&\pcite{aad2023}%
~~& 125.1261(48)~ & $\p{<}\;0.1$ & ---$^*$ & ---\\ %
\botrule
\end{tabular}
\chap{\end{center}}\chapeight{\end{center}}
\label{tab:masses}
}~\\
\\
$^*$~The mass of the Higgs in the Standard Model is a measured rather than a derived quantity. Theoretical considerations placed some broad constraints on the Higgs mass prior to its detection in 2012, but these are not a calculated value in the sense considered here.
\end{table}

Regarding confidence intervals for CASMIR in \tref{tab:masses}, it would be easy to assume that unevaulated higher-order terms such as those in $\mc{O}_b$~\eref{eq:VIII:Ob} and $\mc{O}_e$~\eref{eq:VIII:Oe} attract coefficients of $\ILO{1}$ or less. However, in CASMIR situations frequently arise where $\ILO{10}$ degenerate channels may reinforce one another (e.g.~equivalent loop correction diagrams due to eight species of colored $W$ bosons and one colorless $W$ boson). The confidence intervals for CASMIR in \tref{tab:masses} are therefore evaluated for coefficients in the range $\pm10$. While this is a reasonable precaution in the absence of any exploration of these higher-order terms, it may potentially overestimate the uncertainty associated with the CASMIR values of the calculated parameters, artificially reducing tension. As a check on the values calculated, an upper bound on tension between experiment and the CASMIR expressions to present order is then obtained by taking the CASMIR uncertainty to zero in the calculation of tension. This precaution has negligible effect on the tensions obtained, with all upper bounds on tensions \draftone{for the CASMIR model} in \tref{tab:masses} being $0.2\,\sigma$ or less.

\end{document}